# Direct real-space imaging of sign-reversal superconducting state by scanning Josephson tunneling

Tianheng Wei[1]†, Yanzhao Liu[2]†, Wei Ren[1]†, Zhen Liang[1,5], Ziqiang Wang[3], Jian Wang[1,4,5]*

[1] International Center for Quantum Materials, School of Physics, Peking University; Beijing 100871, China

[2] Quantum Science Center of Guangdong–Hong Kong–Macao Greater Bay Area (Guangdong); Shenzhen 518045, China

[3] Department of Physics, Boston College; Chestnut Hill, MA 02467, USA

[4] Collaborative Innovation Center of Quantum Matter; Beijing 100871, China

[5] Hefei National Laboratory; Hefei 230088, China

†These authors contributed equally to this work

*Corresponding author. Email: jianwangphysics@pku.edu.cn

**Abstract:**

Phase-sensitive experiments based on Josephson tunneling have been considered as a pivotal method to determine the pairing symmetry in high-temperature (high-$T_c$) cuprate superconductors. However, in iron-based superconductors, traditional phase-sensitive experiments face setbacks because of the complexity of pairing symmetry in multiband systems. Here, using scanning Josephson tunneling microscopy (SJTM), we report the half-unit-cell periodic Josephson tunneling modulations in monolayer high-$T_c$ Fe(Te,Se) film epitaxially grown on $SrTiO_3$(001). The Josephson tunneling almost vanishes along the lines connecting next-nearest-neighbor Fe atoms under an *s*-wave superconducting tip. Combined with single-particle tunneling spectra showing the full superconducting gap, our experimental results are consistent with the sign-reversal nodeless *d*-wave pairing symmetry. Our findings provide sub-unit-cell scale superconducting phase information for understanding the mechanism of high-$T_c$ superconductivity. We anticipate our work as a starting point for atomically resolved Josephson interference experiments using SJTM in various unconventional superconductors.

**Main Text:**

The determination of the superconducting pairing symmetry is a crucial step in identifying the underlying mechanism of unconventional high-temperature (high-$T_c$) superconductivity. For high-$T_c$ cuprate superconductors, a series of phase-sensitive experiments (*1–4*) provided significant evidence in favor of the *d*-wave symmetry. These experiments are based on Josephson tunneling and flux quantization, which are directly sensitive to the phase of the superconducting order parameter. Iron-based superconductors (FeSCs), the second family of high-$T_c$ superconductors, possess multiple Fermi surfaces which allow for more pairing symmetry candidates (*5*, *6*). Weakly or moderately doped FeSCs with both electron and hole Fermi pockets are proposed to have $s_{\pm}$ pairing gap function where the superconducting order parameter has opposite signs between electron and hole Fermi pockets (*7–9*). On the other hand, the pairing symmetry remains less understood for heavily doped FeSCs with only one type of Fermi pockets, such as the monolayer FeSe and Fe(Te,Se) films grown on $SrTiO_3$(001) (STO) substrates (*10–19*), which have only electron Fermi pockets (*20*) and exhibit the highest superconducting transition temperatures among FeSCs (*21*, *22*). However, phase-sensitive experiments on FeSCs are relatively rare and only focus on moderately doped FeSCs (*23–25*) due to the difficulties in fabricating high-quality, homogeneous Josephson junctions at macroscopic or mesoscopic scales. Moreover, the complexity of pairing symmetry in multiband systems presents a significant challenge in interpreting traditional Josephson tunneling experiments (*5*, *6*), which lack both spatial and momentum resolution. Therefore, unambiguously identifying the pairing symmetry of unconventional high-$T_c$ superconductors requires a more accessible phase-sensitive experimental approach with high resolution in space or momentum.

Scanning Josephson tunneling microscopy (SJTM) is a cutting-edge technique in scanning tunneling microscopy (STM) experiments (*26–28*), in which a superconducting tip is used to form a spatially tunable microscopic Josephson junction with the superconducting sample. The microscopic Josephson junction in SJTM avoids the inhomogeneity problem present in macroscopic junctions and simultaneously enables subatomic spatial resolution through the manipulation of the tip position. Despite these advantages, the SJTM operates in the dynamical Coulomb blockade regime where the gauge-invariant phase-difference in the Josephson tunneling between the tip and the sample is strongly fluctuating (*26*, *29–31*). However, if there are multiple Josephson tunneling channels, their interference depends on the relative phase-difference between the different channels, which can survive in this regime and provide rudimentary phase-sensitivity for the SJTM (*32*). Thus, SJTM experiments have the potential to provide Josephson interference information in real space to probe the pairing symmetry of unconventional superconductors.

In this study, by using SJTM, we report the half lattice unit cell periodic modulations of the zero-bias Josephson tunneling conductance between an *s*-wave superconducting tip and the monolayer Fe(Te,Se)/STO film, which is a two-dimensional high-$T_c$ FeSC. The Josephson tunneling signal almost vanishes along the lines connecting the next-nearest-neighbor Fe atoms, which are the diagonal mirror lines of the lattice. Together with the fully opened superconducting gap revealed by single-particle tunneling spectra, our experimental results provide the first microscopic real-space interference evidence for the sign-reversal of the superconducting order parameter that is consistent with the nodeless *d*-wave pairing symmetry. Such SJTM experiments are further performed on several superconductors with different pairing symmetries, including the close cousin monolayer FeSe/STO film, bulk Fe-based superconductor Fe(Te,Se), conventional superconductor Pb film on Si(111), and high-$T_c$ cuprate superconductor $Bi_2Sr_2CaCu_2O_{8+x}$. Similar sub-unit-cell Josephson tunneling modulations with vanishing signals along the diagonal mirror lines are observed in monolayer FeSe/STO and $Bi_2Sr_2CaCu_2O_{8+x}$ with *d*-wave pairing symmetry, whereas absent in *s*-wave superconductors (Pb/Si(111) and bulk Fe(Te,Se)). These findings together demonstrate that our SJTM technique is robust and reliable. Our experimental method can be further extended to a broad class of novel superconducting systems, such as other heavily doped iron-based superconductors and cuprates, heavy-fermion superconductors, and moiré superconductors, offering the potential to provide new insights into the pairing mechanism of unconventional superconductors.

**Scanning Josephson tunneling on monolayer Fe(Te,Se)/STO**

The high-quality one-unit-cell (1-UC) thick Fe(Te,Se)/STO film is prepared by molecular beam epitaxy (MBE) (*33*). Figure 1A displays the crystal structure of the 1-UC Fe(Te,Se) film comprising one Fe layer sandwiched between two Te/Se layers. The Fe atoms are arranged on a square lattice with a lattice constant of $a_{\mathrm{Fe}}$ and tetrahedrally coordinated with Te/Se atoms above and below the Fe layer. The top chalcogen layer has a square lattice structure with a lattice constant of $a_{\mathrm{Te/Se}} = \sqrt{2}a_{\mathrm{Fe}}$, which is visible in the STM topography. As shown in the right panel of Fig. 1A, we denote the one-Fe unit cell (grey square) axes as $x$ and $y$, and the two-Fe unit cell (blue square) axes as $a$ and $b$. The corresponding Brillouin zones (BZ) are schematically shown in Fig. 1B. Since the primitive unit cell of the 1-UC Fe(Te,Se)/STO film is the two-Fe unit cell, the folded two-Fe BZ in Fig. 1B is used throughout this work unless otherwise specified. Due to the heavy electron doping, there are only electron Fermi surfaces around the $M$ points in the two-Fe BZ, which is confirmed by our quasiparticle interference experiments (fig. S1). The symmetry elements of the crystal lattice and the Fermi surface structure are marked in Fig. 1, A and B, including the diagonal mirror plane $m_d$ (green dashed line), the vertical mirror plane $m_v$ (yellow dashed line), and the four-fold rotation axis $C_4$ (purple dot). Note that the coordinate origin for these symmetry operations is fixed at the projection of the Te/Se atom onto the Fe plane, and thus these operations are point-group symmetry operations of the non-symmorphic $P4/nmm$ crystal structure. Figure 1C shows a typical STM topography of the 1-UC Fe(Te,Se)/STO with the topmost Te/Se lattice clearly resolved. The sites of different atoms are marked for several unit cells. The Te concentration of the $FeTe_xSe_{1-x}$ film $x \approx 0.3$ is estimated through the thickness of the 2$^{\mathrm{nd}}$ layer (0.58 nm, fig. S2) (*34*). The single-particle tunneling spectra measured with a metallic PtIr tip show the fully gapped superconductivity with the gap energy larger than 10 meV at 4.3 K, and the superconducting gap closing temperature of around 58 K (fig. S2), which are consistent with previous results on the interface-enhanced superconductivity (*13*, *34–36*).

To directly probe the Cooper-pair condensate in the 1-UC Fe(Te,Se)/STO, we perform the SJTM experiments with the superconducting Pb-coated PtIr tip at ultralow temperatures (below 150 mK) as shown in Fig. 1D. As the superconducting tip approaches the sample surface, two hallmarks of fluctuation dominated Cooper-pair tunneling spectroscopy (*26*), the local maximum current ($I_m$) and the zero-bias conductance peak due to Josephson tunneling ($g_J$), emerge in the $I/V$ and $\mathrm{d}I/\mathrm{d}V$ curves, respectively (Fig. 1, E and F). The $I_m$ is proportional to the zero-bias conductance $g_J = \mathrm{d}I/\mathrm{d}V$ ($V = 0$) (fig. S3), and both $I_m$ and $g_J$ are proportional to the square of the Josephson critical current $I_c$ (*26*, *29–31*, *37*). As reported before, $I_m$ and $g_J$ are powerful tools to characterize the local superconductivity information (*27*, *28*, *37*, *38*). To gain more spatial information of the superconductivity, we compare the Josephson tunneling spectra obtained on different sites as shown in Fig. 1G. Strikingly, the Josephson tunneling almost vanishes on all atomic sites but exhibits a large signal at the midpoints of the nearest-neighbor Fe-Fe bonds (also the midpoints between the top and bottom Te/Se sites). We also measure the tunneling spectra in a larger bias range (within ±14 mV) at these four kinds of positions (fig. S4). At different positions, the single-particle tunneling spectra show comparable fully opened superconducting gaps at large bias (fig. S4A), while the Josephson tunneling spectra around zero bias exhibit large variations (fig. S4B). It is therefore safe to rule out the possible influence from changes in the normal state tunneling conductance. Such a huge difference occurs when the tip position changes by a distance (around 0.14 nm) that is much smaller than the superconducting coherence length (around 1.7 nm) (*39*), suggesting a different physical origin other than the inhomogeneous superfluid reported in bulk $FeTe_{0.55}Se_{0.45}$ (*37*).

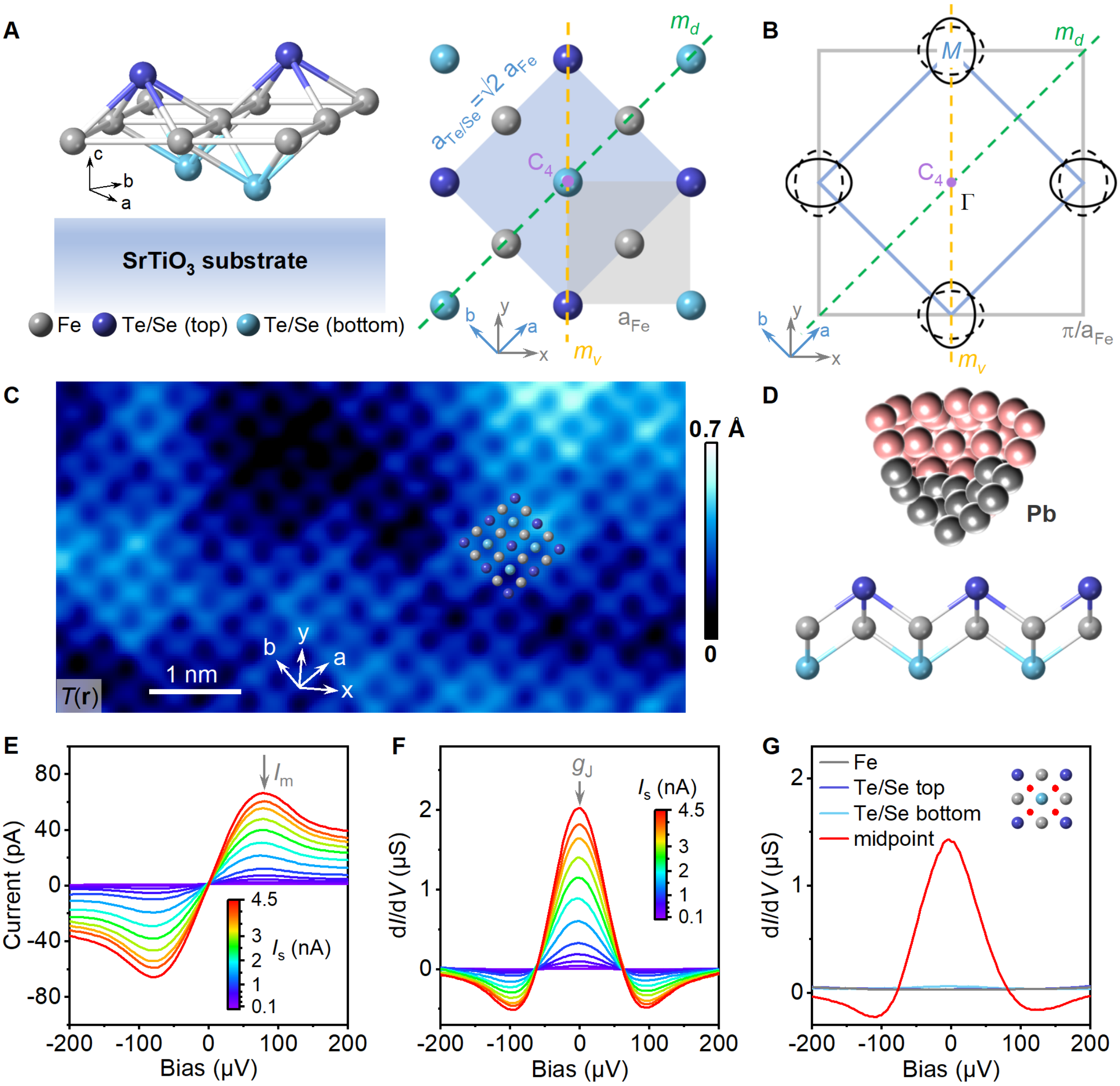

**Fig. 1. Josephson tunneling spectra on the 1-UC Fe(Te,Se)/STO film.** (**A**) The crystal structure of the 1-UC Fe(Te,Se)/STO film. (**B**) Schematic of the Fermi surface of 1-UC Fe(Te,Se)/STO. Grey and blue squares represent one-Fe BZ and two-Fe (folded) BZ, respectively. The original and folded Fermi surfaces are plotted by solid and dashed black lines, respectively. In (A) and (B), the diagonal ($m_d$) and vertical ($m_v$) mirror planes and the $C_4$ axis are marked by green dashed line, yellow dashed line, and purple dot, respectively. The coordinate origin for the symmetry operations is fixed at the projection of the Te/Se atom onto the Fe plane. (**C**) An atomically resolved topography of the 1-UC Fe(Te,Se)/STO film (6.8×3.6 $nm^2$, $V_s$ = -40 mV, $I_s$ = 4 nA). The sites of Fe and Te/Se atoms within several unit cells are marked by colored balls. (**D**) A schematic of the SJTM setup. (**E**) Current-voltage (*I-V*) curves measured with a superconducting tip for different tip-sample distances, in which the local maximum current is marked as $I_m$. (**F**) Differential conductance spectra simultaneously measured with (E). Measurement condition of (E) and (F): $T$ = 95 mK, $V_s$ = 5 mV, $V_{mod}$ = 50 μV. (**G**) Typical Josephson tunneling spectra measured on the different sites ($T$ = 150 mK, $V_s$ = 5 mV, $I_s$ = 4 nA, $V_{mod}$ = 50 μV). The Josephson tunneling almost vanishes on all atomic sites

but has large signal at midpoints of the nearest-neighbor Fe sites, which are schematically shown by red dots in the inset.

## Half lattice periodic modulation of the Josephson tunneling conductance

We now focus on the unusual spatial variation of the Josephson tunneling signal. As shown in Fig. 1G inset, the midpoints of the Fe-Fe bond form a square lattice with axes along *a* and *b* directions and the lattice constant $a_{Te/Se}/2$. Thus, high-spatial-resolution Josephson tunneling spectra (Fig. 2B) are measured along the *b* direction, passing through the midpoints between nearest-neighbor Fe atoms (orange arrow in Fig. 2A). As shown in Fig. 2, B and C, the Josephson tunneling spectra show clear periodic modulations with a period of $a_{Te/Se}/2$. Figure 2, D and E show the modulation of $g_J$ extracted from Fig. 2B and the corresponding Fourier transform. The $g_J$ takes its maximum at midpoints between the nearest-neighbor Fe atoms and almost vanishes at midpoints of nearest-neighbor Fe-Te/Se atoms, exhibiting $a_{Te/Se}/2$ periodic modulation. We also measure the Josephson tunneling spectra along *x* direction, passing through the nearest-neighbor Fe atoms and their midpoints as shown in fig. S5. The $g_J$ shows $a_{Fe}$ periodic modulations, peaks at midpoints and almost vanishes on Fe atomic sites.

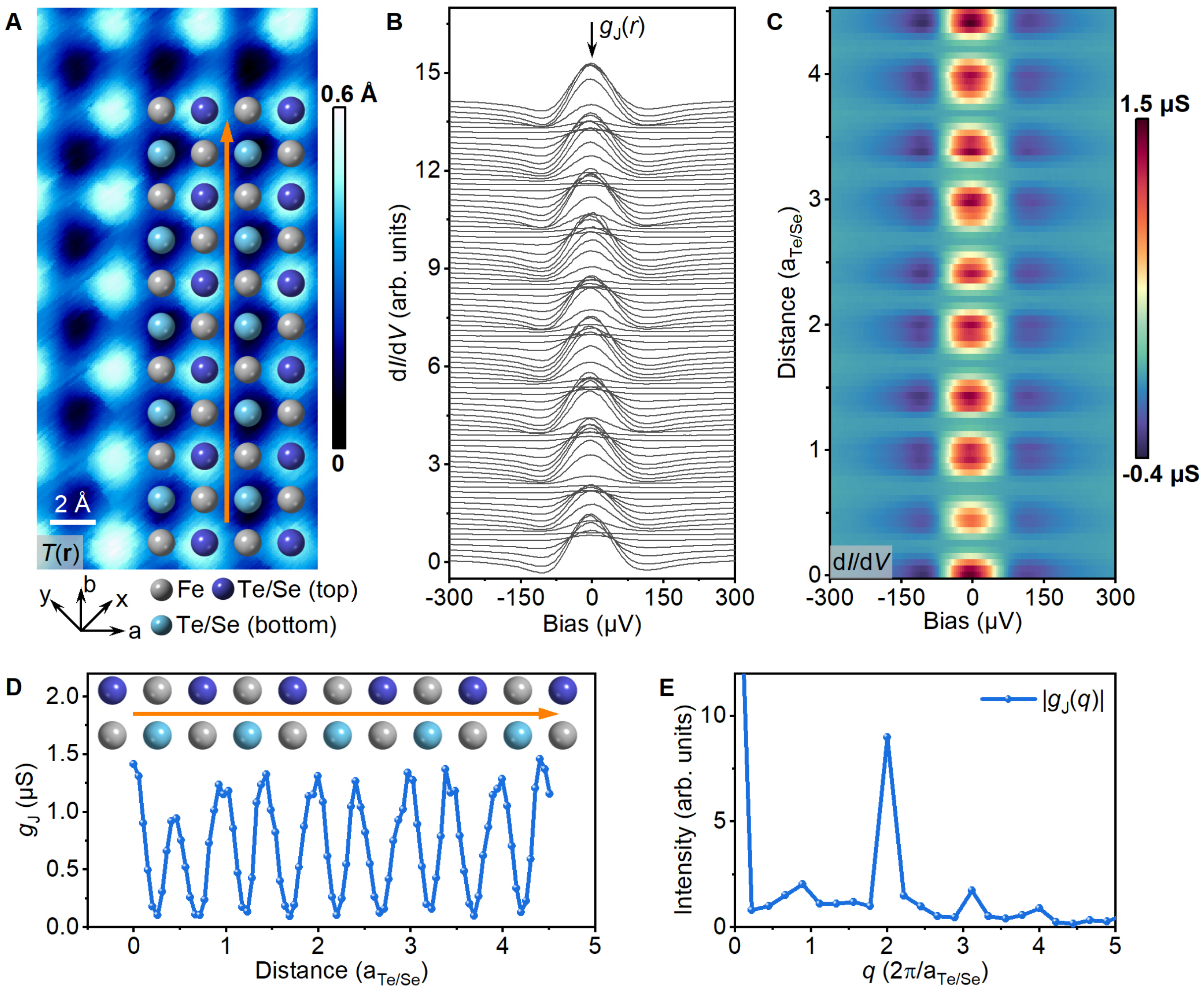

**Fig. 2. Josephson tunneling signal modulation with the period of $a_{Te/Se}/2$ in the 1-UC Fe(Te,Se)/STO film.** (**A**) A topographic image of the 1-UC Fe(Te,Se)/STO film (1.2×2.5 nm$^2$, $V_s = 5$ mV, $I_s = 4$ nA). (**B**) Josephson tunneling spectra measured along the orange arrow in (A) ($T = 70$ mK, $V_s = 5$ mV, $I_s = 4$ nA, $V_{mod} = 50$ μV). The curves are vertically shifted for clarity. The black arrow indicates the Josephson tunneling conductance at zero bias $g_J$. (**C**) The color map of (B), which clearly exhibits the spatially modulated Josephson tunneling signals. (**D**)

The extracted $g_J$ along the arrow in (A). The $g_J(r)$ curve shows spatial modulations with the period of half the Te/Se lattice constant $a_{Te/Se}/2$ and the $g_J$ values are almost zero at the modulation minima. The distances in (C) and (D) are defined relative to the beginning of the arrow in (A). (**E**) The magnitude of the Fourier transform of (D), showing a strong peak at $q = 4\pi/a_{Te/Se}$.

**Imaging the Josephson tunneling conductance map in 1-UC Fe(Te,Se)/STO**

To gain further insights into the spatial modulations of Josephson tunneling, we carry out spectroscopic imaging SJTM measurements on the 1-UC Fe(Te,Se)/STO film. In the region shown in Fig. 3A, we measure the zero-bias conductance of Josephson tunneling $g_J$ at every pixel to obtain the spatial map $g_J(\mathbf{r})$ (Fig. 3B). The Fourier transform of the topography $|T(\mathbf{q})|$ (Fig. 3C) shows distinct Fourier peaks at $\mathbf{Q}_{a,b}$, which are the reciprocal lattice vectors of the topmost Te/Se lattice (two-Fe unit cell). Intriguingly, as shown in Fig. 3D, the dominant modulation wavevectors of $g_J(\mathbf{r})$ are located at $2\mathbf{Q}_{a(b)}$, distinct from the modulations at $\mathbf{Q}_a$ and $\mathbf{Q}_b$ in the atomic topography (Fig. 3C). The different wavevectors between the modulations of $g_J(\mathbf{r})$ and topography $T(\mathbf{r})$ demonstrate that the setup effect in the $g_J(\mathbf{r})$ map does not influence the observed modulation in Josephson tunneling. To further confirm this result and exclude possible influence from the setup effect, we perform the SJTM measurements with different setup conditions and obtain similar $2\mathbf{Q}_{a(b)}$ modulation in the $g_J(\mathbf{r})$ map (fig. S6).

Since the $g_J(\mathbf{r})$ map is obtained simultaneously with the topography $T(\mathbf{r})$, these two maps have pixel-to-pixel correspondence and we can precisely determine the atomic positions in the $g_J(\mathbf{r})$ map. In Fig. 3, A and B, the atomic positions within several unit cells are marked by colored balls according to the topography $T(\mathbf{r})$. The $g_J(\mathbf{r})$ map shows that the Josephson tunneling signal almost vanishes not only above all atomic sites, but also along the lines connecting the next-nearest-neighbor Fe atoms (light yellow lines in Fig. 3B), which coincide and lie in the diagonal mirror plane $m_d$ of the crystal lattice (green lines in Fig. 3, A and B). Such lines of zeroes in the Josephson tunneling signal are also observed under different setup conditions (fig. S6) and using a different microscopic superconducting tip (fig. S7), which demonstrate that they are intrinsic properties of the high-$T_c$ superconducting state in the 1-UC Fe(Te,Se)/STO. The observation of the novel half lattice periodic modulations of the Josephson tunneling with the "zero lines", which has never been reported in previous SJTM experiments (*27*, *28*, *37*, *38*, *40*), indicates the microscopic phase information of the superconductivity in the 1-UC Fe(Te,Se)/STO.

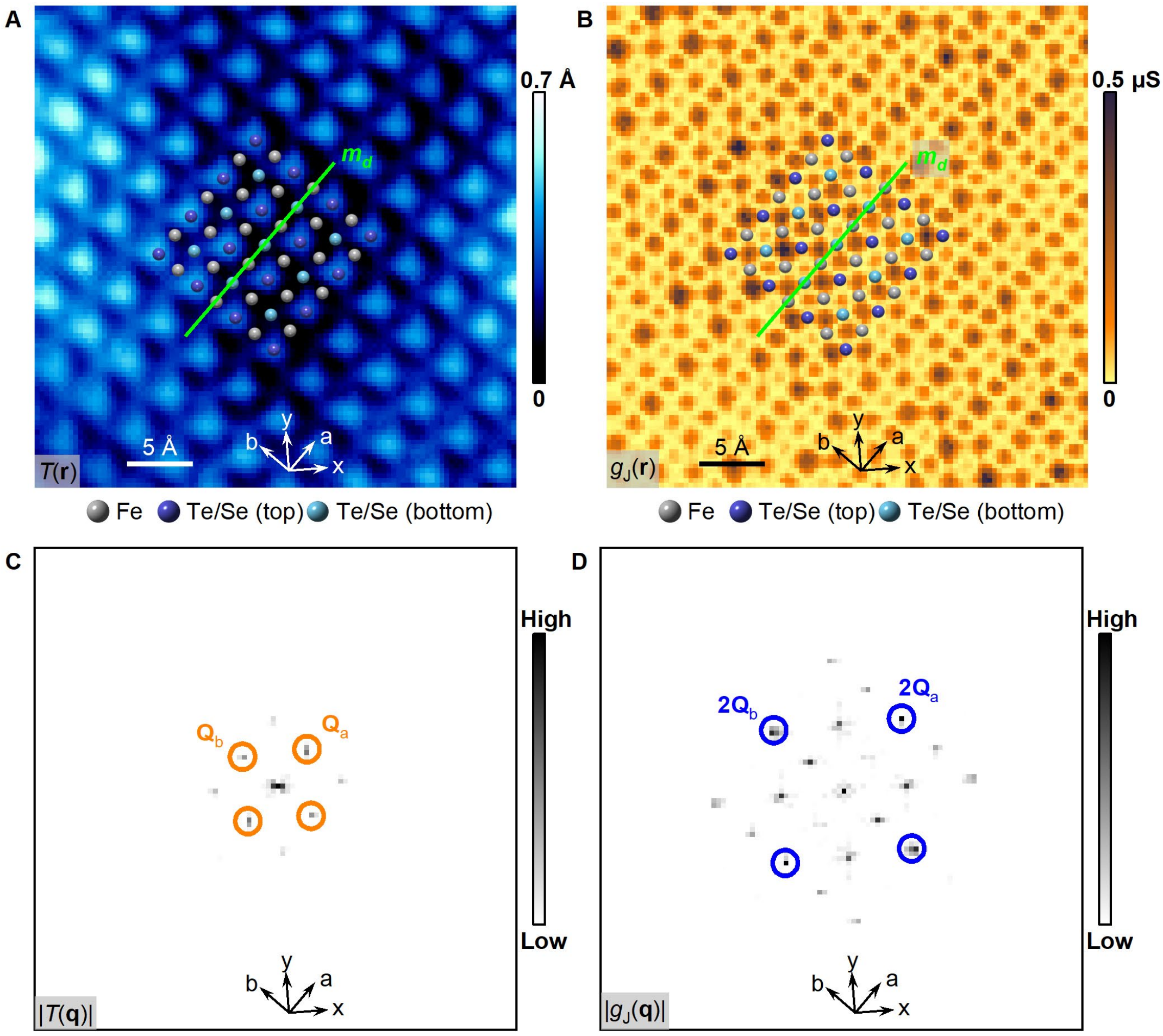

**Fig. 3. Half lattice periodic modulation of the Josephson tunneling signal with zero lines in the 1-UC Fe(Te,Se)/STO film.** (**A**) A topographic image of the 1-UC Fe(Te,Se)/STO film ($3.7\times3.7$ nm$^2$, $V_s$ = 5 mV, $I_s$ = 2 nA). (**B**) The $g_J(\mathbf{r})$ map measured in the same area as in (A), which shows half lattice periodic modulations ($T$ = 95 mK, $V_s$ = 5 mV, $I_s$ = 2 nA, $V_{mod}$ = 50 μV, 816 pixels/nm$^2$). The topography (A) and $g_J(\mathbf{r})$ map (B) are obtained simultaneously, showing pixel-to-pixel correspondence between these two datasets. The sites of Fe and Te/Se atoms within several unit cells are marked by colored balls. The green lines show the diagonal mirror plane $m_d$ of the crystal lattice, where $g_J$ is almost zero. (**C** and **D**) The magnitude of the Fourier transform of (A) and (B), respectively. The modulation wavevector peaks at the Bragg points of the topmost Te/Se lattice ($\mathbf{Q}_a$ and $\mathbf{Q}_b$) are denoted by orange circles in (C). The modulation wavevector peaks at $2\mathbf{Q}_a$ and $2\mathbf{Q}_b$ are denoted by blue circles in (D).

## Discussion

The Josephson tunneling signals $g_J(\mathbf{r})$ and $I_m(\mathbf{r})$ are considered to be approximate estimations of the local superfluid density in previous SJTM studies. However, in these experiments, the superconducting sample and tunneling tip have the same pairing symmetry (*27*, *28*, *38*, *40*). For example, the modulations of Josephson tunneling signals observed as evidence for pair density waves in cuprates (*28*) and 2H-$NbSe_2$ (*38*) superconductors are measured with a *d*-wave superconductor $Bi_2Sr_2CaCu_2O_{8+x}$ tip and an *s*-wave Nb tip, respectively. In this configuration, the reported Josephson tunneling modulations coexist with a large uniform background and have periods of several unit cells, which is fundamentally different from our observations of the half-unit-cell periodic modulations with "zero lines" in the 1-UC Fe(Te,Se)/STO. In contrast, in our measurements, the superfluid density is unlikely to have such large modulations (close to 100%) with a period ($a_{Te/Se}/2 \approx 0.19$ nm) much shorter than the superconducting coherence length (around 1.7 nm) (*39*).

Since the measured $g_J(\mathbf{r})$ is proportional to $I_c^2$, the zero lines in $g_J(\mathbf{r})$ correspond to the vanishing Josephson critical current $I_c = 0$ on such lines, which is possible when the tip and the sample have different pairing symmetries. For the Josephson tunneling between an *s*-wave tip and a superconductor with different pairing symmetry, the Josephson critical current is a coherent superposition resulting from the quantum interference among multiple Josephson tunneling channels (*2*, *41–43*). The relative phases between Josephson tunneling channels are determined by the relative phase structure of the pairing function. To the lowest order in the tunneling amplitude, the Josephson critical current can be expressed as (*44*):

$$I_c(\mathbf{r}) \propto \int \frac{\mathrm{d}\omega}{2\pi} n_F(\omega) \mathrm{Im}\left[F_s^\dagger(\mathbf{r},\mathbf{r};\omega) F_t(\omega)\right] \qquad (1)$$

where $n_F(\omega)$ is the Fermi distribution function, $F_s$ and $F_t$ are anomalous Green's functions of the sample and the tip, respectively. Since the tip is an *s*-wave superconductor, the symmetry property of $I_c$ is determined by that of $F_s$. In the weak coupling limit, $F_s \propto G_0 \hat{\Delta} G_0^T$ where $G_0$ is the normal state Green's function and $\hat{\Delta}$ is the irreducible representation of the superconducting gap function. Since $G_0$ is invariant, $F_s$, and thus $I_c$, transforms according to the pairing matrix $\hat{\Delta}$ of the sample under symmetry operations. Particularly, if the pairing function of the sample is odd under certain symmetry operation $\hat{O}$, the Josephson critical current transforms as $I_c(\hat{O}\mathbf{r}) = -I_c(\mathbf{r})$, enabling the symmetry-protected zero Josephson critical current [$I_c(\mathbf{r}) = 0$] at the high symmetry positions $\hat{O}\mathbf{r} = \mathbf{r}$.

In our experiments on the 1-UC Fe(Te,Se)/STO, the zero lines of the Josephson tunneling conductance $g_J$ are observed when the SJTM tip is located in the diagonal mirror plane $m_d$ of the lattice (Fig. 3B). This strongly indicates that under the diagonal mirror reflections ($\hat{\sigma}_d$), the pairing function is odd, and the Josephson critical current changes sign, i.e., $I_c(\hat{\sigma}_d\mathbf{r}) = -I_c(\mathbf{r})$, enforcing the symmetry-protected zero lines of Josephson tunneling along the diagonal mirror lines of the lattice ($\hat{\sigma}_d\mathbf{r} = \mathbf{r}$) connecting the next-nearest-neighbor Fe atoms (Fig. 4A). Moreover, $g_J$ shown in Fig. 3B is maximized at the midpoints of the nearest-neighbor Fe-Fe bond, which are invariant under vertical mirror reflection $\hat{\sigma}_v$, suggesting the pairing function is even under $\hat{\sigma}_v$. Together with the full superconducting gap revealed by single-particle tunneling spectra (fig. S2), these symmetry properties are fully consistent with the nodeless *d*-wave pairing symmetry (Fig. 4B) theoretically proposed for both 1-UC Fe(Te,Se)/STO films and heavily electron-doped bulk FeSCs (*18*, *19*, *45*, *46*).

Previous theoretical studies have provided the complete classification of superconducting pairing irreducible representations and their symmetry properties at the *M* point in the two-Fe BZ for the non-symmorphic space group of monolayer FeSe (*47*, *48*). The nodeless *d*-wave pairing state is identified exactly with the $B_{2g}$ pairing sector (*33*), which is odd under $\hat{\sigma}_d$ and even under $\hat{\sigma}_v$ with respect to the mirror planes $m_d$ and $m_v$ defined in Fig. 1A. In the *d*-wave pairing state (Fig. 4B), the two Fermi pockets hosting different pairing phases cross at four points on the zone boundary, raising the possibility of emerging nodes. However, theoretical studies have shown that the *d*-wave pairing state remains nodeless if the interband spin-orbit coupling (SOC) is smaller than the superconducting gap energy (*18*, *48–50*). Experimentally, no mixing of the two Fermi pockets was observed in 1-UC FeSe/STO film with a 5 meV instrument resolution (*51*), suggesting the interband SOC in this class of superconducting films is smaller than the gap energy (around 10 meV). Furthermore, our SJTM results indicate that the plain *s*-wave (Fig. 4C) or extended $s_\pm$-wave (Fig. 4D) pairing states are highly unlikely since their pairing functions ($A_{1g}$ type) are even under symmetry operations and thus do not support symmetry-protected zero Josephson tunneling lines in real space (see Supplementary Materials (*33*) for detailed symmetry analysis and calculations).

The crucial Josephson tunneling interference is schematically illustrated in Fig. 4A. When the tip is located in the diagonal mirror plane $m_d$ of the 1-UC Fe(Te,Se)/STO, the microscopic junction is invariant under $\hat{\sigma}_d$, which ensures that the Josephson tunneling from the Cooper pairs from the red pocket with pairing phase $\varphi_s$ interfere destructively with those from the blue pocket with pairing phase $\varphi_s+\pi$, resulting in the zero lines in the total Josephson tunneling along the diagonal mirror lines connecting the next-nearest-neighbor Fe atoms. To help visualize the SJTM patterns, we calculate the spatial modulation patterns of the SJTM using approximate Wannier functions in the limit of negligible interband SOC and interband pairing (*33*). Figure 4E shows the calculated $I_c(\mathbf{r})$ patterns, exhibiting the lattice periodicity and the sub-unit-cell *d*-wave symmetry. The corresponding $I_c^2(\mathbf{r})$ map is shown in Fig. 4F, which is proportional to the Josephson tunneling conductance $g_J(\mathbf{r})$. Along the diagonal mirror lines of the lattice connecting the next-nearest-neighbor Fe atoms, the $I_c^2$ is zero, which is in very good agreement with our experimentally measured $g_J(\mathbf{r})$ map shown in Fig. 3B and further supports our observation of a nodeless *d*-wave pairing symmetry in the 1-UC Fe(Te,Se)/STO.

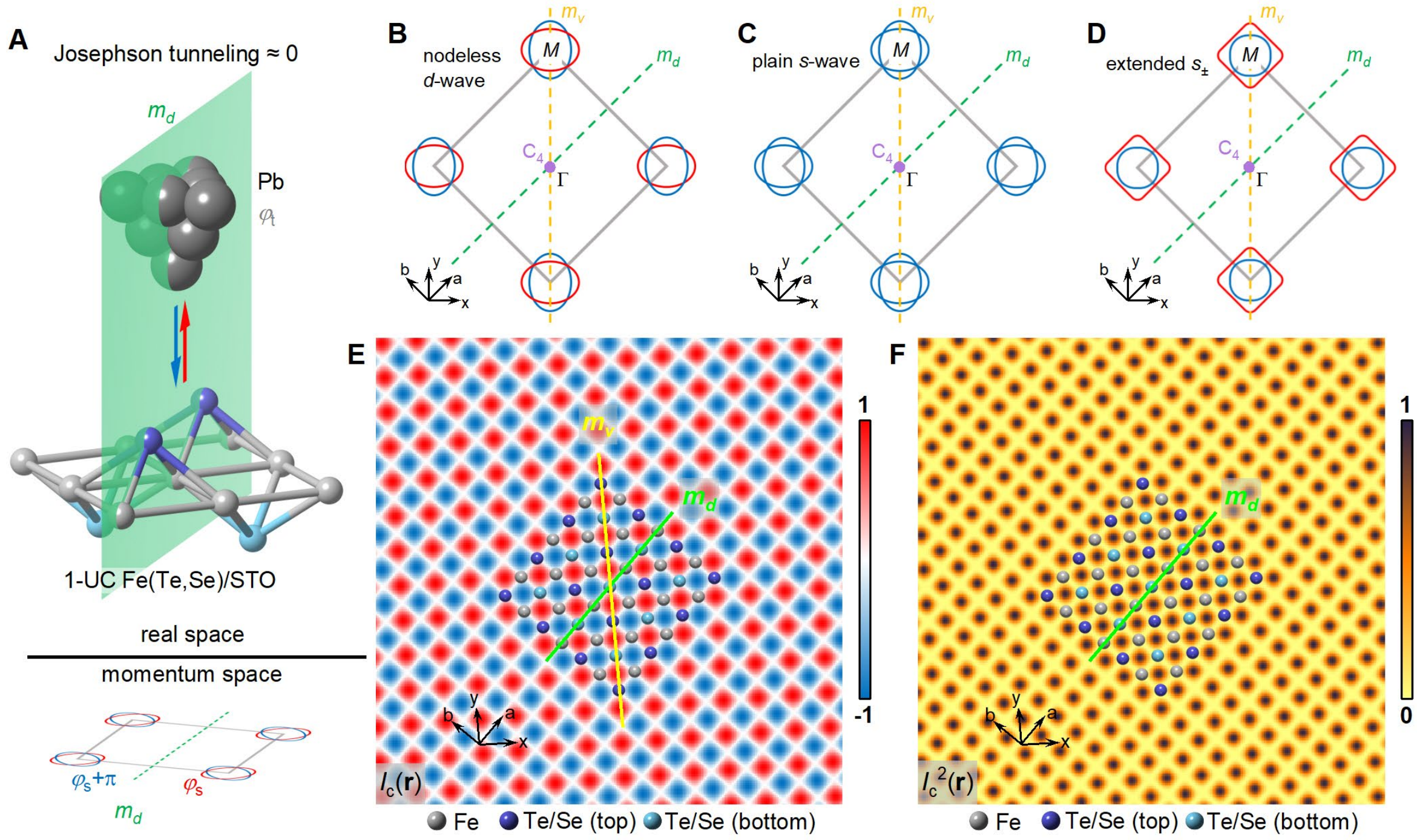

**Fig. 4. Josephson tunneling interference and nodeless *d*-wave pairing symmetry.** (**A**) Schematics for the interference of Josephson tunneling when the tip is located in the diagonal mirror plane. The superconducting phase of the Pb tip is $\varphi_t$. The superconducting phases on the two Fermi pockets (denoted by red and blue ellipses) of the 1-UC Fe(Te,Se)/STO film are $\varphi_s$ and $\varphi_s+\pi$, respectively. The diagonal mirror plane is denoted by the green plane in real space and the green dashed line in momentum space. The red/blue arrow in real space panel (top) represents the Josephson tunneling associated with the red/blue Fermi pocket in momentum space. (**B**-**D**) Schematics for the nodeless *d*-wave, plain *s*-wave, and extended $s_\pm$ pairing states in the two-Fe BZ. Different colors denote the two Fermi surface pockets with opposite signs of the pairing gap functions. The symmetry elements in **k**-space, including the mirror planes $m_d$, $m_v$ and the $C_4$ axis, are marked by green dashed line, yellow dashed line and purple dot, respectively. (**E**) The calculated Josephson critical current $I_c(\mathbf{r})$ between an *s*-wave superconducting tip and the monolayer Fe(Te,Se)/STO with nodeless *d*-wave pairing symmetry (*33*). The $I_c(\mathbf{r})$ map is normalized with respect to its magnitude over the midpoints of the nearest-neighbor Fe atoms. The green and yellow lines show two mirror planes $m_d$ and $m_v$ of the crystal lattice. (**F**) Normalized $I_c^2(\mathbf{r})$ map calculated from (E), which is consistent with the experimental results of the Josephson tunneling conductance map $g_J(\mathbf{r})$ shown in Fig. 3B.

To further support our findings and to test the robustness and reliability of the Josephson tunneling interference technique, we also perform SJTM experiments on various superconductors for comparison, including the monolayer FeSe/STO, the bulk Fe-based superconductor Fe(Te,Se), the conventional superconductor Pb film grown on Si(111), and the high-$T_c$ cuprate $Bi_2Sr_2CaCu_2O_{8+x}$. The Josephson tunneling conductance maps $g_J(\mathbf{r})$ as well as the corresponding topography are summarized in Fig. 5, and more detailed results are shown in figs. S8 to S11. In the closely related monolayer FeSe/STO, we observe Josephson tunneling modulations with zero lines (Fig. 5, A and E), which are indeed similar to the results obtained in the monolayer Fe(Te,Se)/STO, suggesting that the nodeless *d*-wave pairing is not induced by Te substitution, but rather a universal pairing symmetry for the heavily electron-doped monolayer Fe-based superconductors. In the bulk Fe(Te,Se), which exhibits both electron and hole Fermi pockets and is widely recognized as an *s*-wave superconductor with an $s_{\pm}$ pairing gap function, the Josephson tunneling conductance is relatively uniform and shows spatial inhomogeneity over length scales much larger than the atomic spacing (Fig. 5, B and F). This is consistent with the previous report (*37*), but totally different than our observations of sub-unit-cell modulations in the monolayer Fe(Te,Se) or FeSe on STO substrate. In the Pb film grown on Si(111), which is a conventional *s*-wave superconductor, the Josephson tunneling signal is uniform (Fig. 5, C and G). Strikingly, the SJTM experiments on the BiO surface of the cuprate high-$T_c$ superconductor $Bi_2Sr_2CaCu_2O_{8+x}$ reveal similar Josephson tunneling modulations to those in the monolayer Fe(Te,Se) or FeSe on STO. As shown in Fig. 5, D and H, the Josephson tunneling signal is maximized above the O sites between the nearest-neighbor Cu sites in the $CuO_2$ plane (the visualized atoms are Bi atoms, which reside directly above the Cu atoms in the $CuO_2$ plane underneath), but almost vanishes above the lines connecting the next-nearest-neighbor Cu sites, which are the diagonal mirror lines of the lattice (see fig. S11 for more details). This agrees with the $d_{x^2-y^2}$-wave pairing symmetry of cuprate superconductors. Together, these SJTM results on superconductors of different pairing symmetries strongly support our discovery of the nodeless *d*-wave pairing in heavily electron-doped monolayer FeSCs, a pairing symmetry that is different from the bulk Fe(Te,Se) superconductors. Furthermore, they demonstrate that the scanning Josephson tunneling interference technique is powerful enough to distinguish between *d*-wave pairing and *s*-wave pairing at the atomic scale, and thus a useful technique for directly probing the pairing symmetries of unconventional superconductors. In particular, the SJTM technique does not require the fabrication of complicated structures and is well suited for ultrathin superconductors, for which the conventional phase-sensitive experiments, especially those relying on electrical contacts on the side surfaces, are challenging to perform.

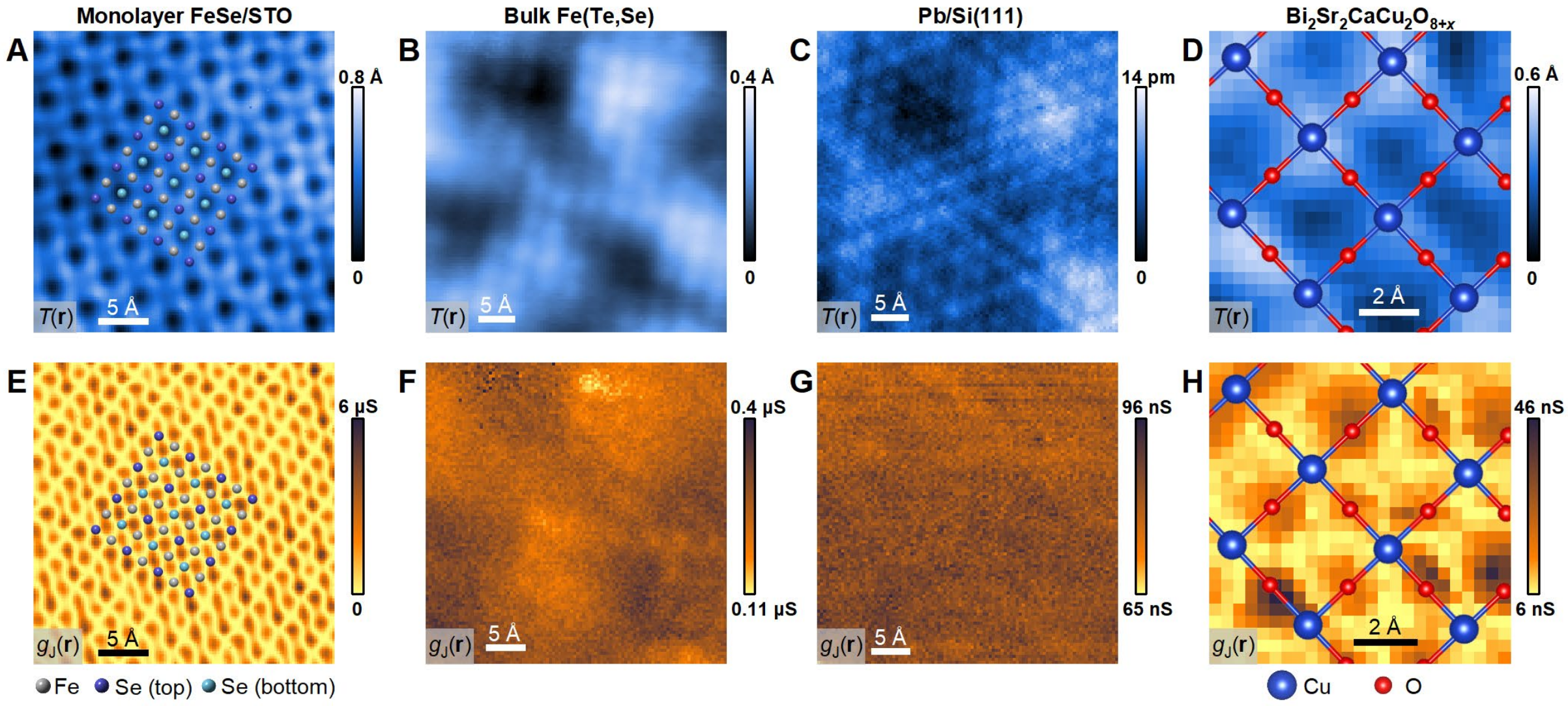

**Fig. 5. SJTM results on various superconductors.** (**A-D**) Atomically resolved topographic images of monolayer FeSe/STO, bulk Fe(Te,Se), Pb/Si(111), and $Bi_2Sr_2CaCu_2O_{8+x}$, respectively. (**E-H**) The Josephson tunneling conductance $g_J(\mathbf{r})$ maps measured in the corresponding areas shown in (A-D). The atoms are marked by colored balls in (A,D,E,H). The $g_J(\mathbf{r})$ maps show sub-unit-cell modulations in monolayer FeSe/STO and $Bi_2Sr_2CaCu_2O_{8+x}$ that are absent in bulk Fe(Te,Se) and Pb/Si(111). Measurement conditions: (A and E) 3.0×3.0 nm$^2$, $T$ = 50 mK, $V_s$ = 5 mV, $I_s$ = 6 nA, $V_{mod}$ = 50 µV; (B and F) 4.0×4.0 nm$^2$, $T$ = 48 mK, $V_s$ = -8 mV, $I_s$ = 20 nA, $V_{mod}$ = 50 µV; (C and G) 4.0×4.0 nm$^2$, $T$ = 49 mK, $V_s$ = 5 mV, $I_s$ = 8 nA, $V_{mod}$ = 50 µV; (D and H) 1.0×1.0 nm$^2$, $T$ = 50 mK, $V_s$ = -5 mV, $I_s$ = 8 nA, $V_{mod}$ = 50 µV.

To conclude, we report the discovery of half lattice periodic modulations of the Josephson tunneling conductance between an *s*-wave superconducting tip and the monolayer Fe(Te,Se)/STO film. The Josephson tunneling modulation is maximized at the midpoints of the nearest-neighbor Fe atoms and almost vanishes along the next-nearest-neighbor Fe-Fe lines. Combined with single-particle tunneling spectra showing the fully opened superconducting gap, our experimental results are most compatible with a nodeless *d*-wave pairing symmetry for the monolayer Fe(Te,Se)/STO. These unusual modulations could be the first visualization of the *d*-wave superconductivity at the atomic scale through Josephson interference experiments. Similar sub-unit-cell modulations with zero lines are further observed in monolayer FeSe/STO and cuprate $Bi_2Sr_2CaCu_2O_{8+x}$, but are absent in bulk Fe(Te,Se) and Pb/Si(111), further supporting our observation and pairing symmetry explanation. Our findings provide precise microscopic evidence for the destructive quantum interference from antiphase superconducting condensates directly in real space within the lattice unit cell and offer new insights into the unconventional superconductors. Our work may also open a new path to study the pairing symmetry of unconventional superconductivity through Josephson interference at the sub-unit-cell scale using the SJTM technique.

**Acknowledgments:**

**Funding:** This work was financially supported by

National Natural Science Foundation of China No. 12488201 (J.W.)

National Natural Science Foundation of China No. 12404215 (Y.L.)

Quantum Science and Technology-National Science and Technology Major Project 2021ZD0302403 (J.W.)

Guangdong Provincial Quantum Science Strategic Initiative GDZX2401001 (Y.L.)

Guangdong Provincial Quantum Science Strategic Initiative GDZX2401009 (Y.L.)

U.S. Department of Energy, Basic Energy Sciences DE-FG02-99ER45747 (Z.W.)

**Author contributions:** J.W. conceived and supervised the research. T.W., W.R. and Y.L. grew the sample and carried out the experiments. T.W., Y.L., Z.L. and J.W. analyzed the experimental data. T.W., Y.L., Z.W. and J.W. wrote the manuscript.

**Competing interests:** The authors declare no competing interests.

**Data and materials availability:** All data are available in the main text or the supplementary materials.

**Supplementary Materials**

Materials and Methods

Supplementary Text

Figs. S1 to S18

References (*52–61*)

## Materials and Methods

Our experiments were performed on a commercial USM-1600 ultra-high vacuum MBE-STM combined system. The Fe(Te,Se) film was grown as the following recipe: The Nb-doped $SrTiO_3$(001) (wt 0.7%) substrate was annealed at 1000°C in the MBE chamber for 1 hour and then cooled down to 330°C for film growth. High purity Fe(99.995%), Te(99.9999%) and Se(99.999%) were co-deposited on the substrate from Knudsen cells for 9 minutes. Then the as-grown film was annealed at 400°C for 17 hours. The thickness of the 1-UC $FeTe_xSe_{1-x}$ film is around 0.58 nm (Fig. S2), corresponding to the composition x ≈ 0.3 (*34*). Because the monolayer Fe(Te,Se)/STO has a relatively large surface corrugation, which is probably due to the roughness of the substrate, the brighter atoms cannot simply be assigned to Te and the darker to Se (*52*). Thus, the Te/Se ratio cannot be determined by counting Te/Se directly from the STM topography (*53*).

After annealing, the films were transported to the *in situ* low temperature STM kept at 4.3 K. The SJTM measurements were performed with a superconducting Pb-coated PtIr tip at ultralow temperatures (below 150 mK). The STM/S measurements at 4.3 K and above (shown in Fig. S2) were performed with a polycrystalline PtIr tip. The STS spectra were acquired by standard lock-in technique at 983 Hz.

## Supplementary Text

Quasiparticle interference (QPI) experiments

The QPI measurements are performed with a PtIr tip at $T$ = 4.3 K. As shown in Fig. S1, the $\mathbf{q}_1$ ring structure in Fig. S1B-E corresponds to the intra-pocket scattering and the energy dispersion of $\mathbf{q}_1$ (Fig. S1F) is consistent with the band dispersion of an electron band, demonstrating the existence of electron pockets at the Fermi level. On the other hand, the $\mathbf{q}_4$ scattering (green dashed arrow in Fig. S1A,B), which connects the electron pocket and a would-be hole pocket, is absent in the QPI patterns. This suggests the absence of the hole pocket around the Γ point at the Fermi level.

Symmetry analyses on SJTM results

Based on the microscopic theory for the Josephson tunneling in continuum space (*44*), we provide symmetry analyses on the SJTM results. The experimentally measured Josephson tunneling conductance $g_J$ is proportional to $I_c^2$, where $I_c$ is the Josephson critical current of the microscopic junction formed by the sample and the superconducting tip. To the lowest order in the tunneling amplitude, $I_c$ can be expressed as:

$$I_c(\mathbf{r}) \propto \int \frac{d\omega}{2\pi} n_F(\omega) \mathrm{Im}\left[F_s^\dagger(\mathbf{r},\mathbf{r};\omega)F_t(\omega)\right]$$

where $n_F(\omega)$ is the Fermi distribution function, $F_s$ and $F_t$ are anomalous Green's functions of the sample and the tip, respectively. Since the tip is an *s*-wave superconductor, the symmetry property of $I_c$ is determined by that of $F_s$ and thus the pairing symmetry of the superconducting sample. We can discuss how $I_c(\mathbf{r})$ transforms under the symmetry operations $\hat{O}$ in the three prominent unconventional superconductors with different superconducting pairing gap functions.

First, for the cuprate superconductors (Fig. S12A), there is only one relevant band with $d_{x^2-y^2}$-wave pairing symmetry. The pairing function is odd under $\hat{C}_4$, $\hat{\sigma}_d$ and even under $\hat{\sigma}_v$. Therefore, we have $I_c(\hat{C}_4\mathbf{r}) = -I_c(\mathbf{r})$, $I_c(\hat{\sigma}_d\mathbf{r}) = -I_c(\mathbf{r})$, and $I_c(\hat{\sigma}_v\mathbf{r}) = I_c(\mathbf{r})$. Given these symmetry constraints, at any position $\mathbf{r}_z$ of the tunneling tip above the next-nearest-neighbor Cu-Cu line, which is in the diagonal mirror plane, we have $\mathbf{r}_z = \hat{\sigma}_d\mathbf{r}_z$ and $I_c(\mathbf{r}_z) = I_c(\hat{\sigma}_d\mathbf{r}_z) =$ $-I_c(\mathbf{r}_z)$. This requires $I_c(\mathbf{r}_z) = 0$ by symmetry, which is consistent with our observations on bulk cuprate $Bi_2Sr_2CaCu_2O_{8+x}$ shown in Fig. S11 and the calculations in Ref. (*44*).

Next, we consider the current study of the monolayer Fe(Te,Se)/STO. In this case, there are two bands derived from the Fe 3*d* orbitals crossing the Fermi level, as shown in Fig. S12B in the one-Fe Brillouin zone (BZ) and in Fig. S12C in the two-Fe BZ by different colors. These two bands are dominated by the *xz*/*xy* and *yz*/*xy* orbitals, respectively (*5*). Independent of whether we discuss in the one-Fe or the two-Fe BZ, we can consistently define the "diagonal" as the Fe-Fe diagonal ($k_x = k_y$ direction). Both the diagonal mirror reflection $\hat{\sigma}_d$: $(x, y) \rightarrow (y, x)$ and the four-fold rotation $\hat{C}_4$: $(x, y) \rightarrow (-y, x)$ exchange these two orbitals/bands, while the $\hat{\sigma}_v$: $(x, y) \rightarrow (-x, y)$ does not. This allows the $\hat{\sigma}_d$-odd orbital-based nodeless *d*-wave pairing order parameter symmetry, where the gap functions on the two Fermi pockets have opposite signs under $\hat{C}_4$ rotation. This is the pairing symmetry observed by our SJTM experiments. These symmetry properties ensure that the Josephson critical current transforms according to: $I_c(\hat{C}_4\mathbf{r}) = -I_c(\mathbf{r})$, $I_c(\hat{\sigma}_d\mathbf{r}) = -I_c(\mathbf{r})$, $I_c(\hat{\sigma}_v\mathbf{r}) = I_c(\mathbf{r})$. As a result, at any position $\mathbf{r}_z$ of the tunneling tip above the next-nearest-neighbor Fe-Fe lines that are in the diagonal mirror plane, the symmetry requires $\mathbf{r}_z = \hat{\sigma}_d\mathbf{r}_z$ and $I_c(\mathbf{r}_z) = I_c(\hat{\sigma}_d\mathbf{r}_z) = -I_c(\mathbf{r}_z)$ and hence $I_c(\mathbf{r}_z) = 0$. This is precisely consistent with our observations in the monolayer Fe(Te,Se)/STO. We note that the superconducting gap is only defined around the two Fermi pockets but not everywhere in the BZ. The *d*-wave symmetry comes from the orbital degrees of freedom under mirror reflection or four-fold rotation. Furthermore, we note that the absence of gap nodes in the orbital-based nodeless *d*-wave scenario makes the zero lines for the Josephson tunneling signal more prominent compared to the case of **k**-space nodal *d*-wave pairing in the cuprates that tends to be influenced by other tunneling processes due to the residual single-particle states near the gap nodes.

Finally, we discuss the $s_\pm$ pairing scenario as shown in Fig. S12D-F. For the usual $s_\pm$ pairing (Fig. S12D in the one-Fe BZ and Fig. S12E in the two-Fe BZ) commonly considered for bulk Fe-based superconductors, the superconducting gap function has opposite signs on the hole pocket around Γ (red) and the electron pockets near the BZ boundary (blue). For the extended $s_\pm$ pairing (Fig. S12F), an alternative scenario for the monolayer Fe(Te,Se)/STO, the pairing gap function has opposite signs on the *xy* pocket (red) and the *xz*/*yz* pocket (blue) (*54*). In both cases of the $s_\pm$ pairing scenario, the pairing order parameter symmetry is *s*-wave, and the Fermi pockets with opposite signs of the gap function are not related by any symmetry operation. Under the symmetry operations $\hat{O}$, we have: $I_c(\hat{C}_4\mathbf{r}) = I_c(\mathbf{r})$, $I_c(\hat{\sigma}_d\mathbf{r}) = I_c(\mathbf{r})$ and $I_c(\hat{\sigma}_v\mathbf{r}) = I_c(\mathbf{r})$. Therefore, there is no symmetry-protected zero Josephson tunneling positions in real space. As a result, the $I_c(\mathbf{r})$ is expected to be relatively uniform in real space, which is consistent with our experimental observations in bulk Fe(Te,Se) with $s_\pm$ pairing shown in Fig. S9, but completely different and cannot describe our SJTM observations in monolayer Fe(Te,Se)/STO (Fig. 3, Fig. S6 and Fig. S7).

Next, we turn to discuss the corresponding real space pattern under different pairing symmetries and compare to our SJTM observations in monolayer Fe(Te,Se)/STO. Because of the translation symmetry, $I_c(\mathbf{r})$ can be expressed as:

$$I_c(\mathbf{r}) = \sum_{m,n} a_{m,n} \cos[(m\mathbf{k}_a + n\mathbf{k}_b) \cdot \mathbf{r}] + b_{m,n} \sin[(m\mathbf{k}_a + n\mathbf{k}_b) \cdot \mathbf{r}]$$

where $m$, $n$ are integers, $\mathbf{k}_a$, $\mathbf{k}_b$ are reciprocal lattice vectors along $a$ and $b$ directions ($|\mathbf{k}_{a,b}| = 2\pi/a_{Te/Se}$). For the nodeless *d*-wave pairing symmetry (Fig. S12, B and C), $I_c(\hat{C}_4\mathbf{r}) = -I_c(\mathbf{r})$ requires $a_{m,n} = -a_{-n,m}$, $a_{0,0} = 0$, $b_{m,n} = 0$. And $I_c(\hat{\sigma}_d\mathbf{r}) = -I_c(\mathbf{r})$ requires $a_{m,n} = -a_{m,-n} = -a_{-m,n}$ thus $a_{m,0} = a_{0,n} = 0$. Therefore, $I_c(\mathbf{r})$ can be expressed up to its lowest-harmonic terms as:

$$I_c(\mathbf{r}) = 2a_{1,1}[\cos(\mathbf{k}_x \cdot \mathbf{r}) - \cos(\mathbf{k}_y \cdot \mathbf{r})] + \cdots$$

where $\mathbf{k}_{y,x} = \mathbf{k}_a \pm \mathbf{k}_b$ are the reciprocal lattice vectors of the Fe lattice along $y$ and $x$ directions ($|\mathbf{k}_{x,y}| = 2\pi/a_{Fe}$). Since the finite size of the tip will effectively smooth the observed real space map and largely suppress the higher-harmonic terms, we only consider the non-zero lowest-harmonic terms and plot the $I_c(\mathbf{r})$ and $I_c^2(\mathbf{r})$ maps along with the corresponding lattice positions in Fig. S13, A and B. The $I_c^2(\mathbf{r})$ map (Fig. S13B) is in very good agreement with our experimentally measured $g_J$ map shown in Fig. 3B and reproduced in Fig. S13C.

Following the symmetry analysis above, we discuss the $I_c(\mathbf{r})$ map for $s_\pm$ pairing. In this case, $I_c(\mathbf{r})$ can be expressed up to its second-lowest-harmonic terms as:

$$I_c(\mathbf{r}) = a_{0,0} + 2a_{1,0}[\cos(\mathbf{k}_a \cdot \mathbf{r}) + \cos(\mathbf{k}_b \cdot \mathbf{r})] + \cdots$$

The spatial pattern of $I_c(\mathbf{r})$ will most likely show a relatively uniform distribution because the tunneling channels from +Δ and -Δ bands are not expected to cancel out since there is no symmetry enforcing the cancelation, resulting in a uniform component $a_{0,0}$. This is clearly inconsistent with our experimental SJTM data in monolayer Fe(Te,Se)/STO (Fig. 3, Fig. S6 and Fig. S7).

<u>Discussions on irreducible pairing functions based on the non-symmorphic *P*4/*nmm* space group of the crystal structure with two-Fe UC</u>

For the identification of the properties of the pairing function or order parameter, the non-symmorphic symmetry and the 2-Fe unit cell have crucial implications, leading to two-component pairing functions. The complete classification of superconducting pairing irreducible representations and their symmetry properties have been developed in a symmetry-adapted low-energy *M*-point little-group basis in the 2-Fe unit cell for monolayer FeSe (*48*). In our experiments, the single-particle tunneling spectra of the monolayer films reveal a full superconducting gap (Figs. S2 and S4), which constrains the possible pairing states. Furthermore, because the spin-triplet pairing states do not contribute to the leading-order Josephson tunneling with the spin-singlet *s*-wave superconducting tip we use, the spin-triplet pairing, including the nodeless helical *p*-wave state, can be ignored. Of all the symmetry-derived irreducible pairing states near the *M* point (*48*), only the spin-singlet $A_{1g}$ (*s*-wave) and the $B_{2g}$ (*d*-wave) have nodeless gap on the Fermi surfaces and can be consistent with our experiments. Using the notations of Ref. (*47*, *48*), the two electronically relevant *M*-point representations are *M*1 and *M*3, forming the basis $\psi_{M,s} = \left(c^X_{M1,s}, c^X_{M3,s}, c^Y_{M1,s}, c^Y_{M3,s}\right)^T$, where $s$ denotes the spin, $X$ and $Y$ denote the two pockets around the *M* point. The pairing matrices for the two states are

$$\hat{\Delta}_{A_{1g}} = \begin{pmatrix} \Delta_1 & 0 & 0 & 0 \\ 0 & \Delta_3 & 0 & 0 \\ 0 & 0 & \Delta_1 & 0 \\ 0 & 0 & 0 & \Delta_3 \end{pmatrix} \otimes (is_2), \qquad \hat{\Delta}_{B_{2g}} = \begin{pmatrix} \Delta_1 & 0 & 0 & 0 \\ 0 & \Delta_3 & 0 & 0 \\ 0 & 0 & -\Delta_1 & 0 \\ 0 & 0 & 0 & -\Delta_3 \end{pmatrix} \otimes (is_2)$$

where $s_2$ is the Pauli matrix in spin space, $\Delta_1$ and $\Delta_3$ are independent order parameters. They are both intraband pairing states, which is consistent with the absence of mixing of the Fermi surfaces in previous ARPES experiments on monolayer $FeSe/SrTiO_3$ and a negligible interband spin-orbit coupling (SOC) (*48*, *51*). We next determine explicitly by symmetry analysis which one of these describes our Josephson tunneling data.

In our SJTM measurements, the Josephson tunneling conductance $g_J$ vanishes along the diagonal mirror (Se-Fe-Se line, $m_d$ in Fig. 1A), suggesting the superconducting order parameter is odd under the diagonal mirror reflection $\hat{\sigma}_d$. The $g_J$ is maximized at the midpoints between nearest-neighbor Fe atoms, which are in the vertical mirror plane ($m_v$ in Fig. 1A), suggesting the superconducting order parameter is even under the vertical mirror reflection $\hat{\sigma}_v$. The four-fold rotation is a combination of the two mirror reflections: $\hat{C}_4 = \hat{\sigma}_d\hat{\sigma}_v$, under which the superconducting order parameter is odd. In the above basis at *M* point, the symmetry operations are (*48*):

$$U(\hat{\sigma}_v) = \begin{pmatrix} -1 & 0 & 0 & 0 \\ 0 & 1 & 0 & 0 \\ 0 & 0 & -1 & 0 \\ 0 & 0 & 0 & -1 \end{pmatrix} \otimes (-is_1), \qquad U(\hat{\sigma}_d) = \begin{pmatrix} 0 & 0 & 1 & 0 \\ 0 & 0 & 0 & 1 \\ 1 & 0 & 0 & 0 \\ 0 & 1 & 0 & 0 \end{pmatrix} \otimes \left(-i\frac{s_1 - s_2}{\sqrt{2}}\right)$$

where $s_1$ and $s_2$ are the Pauli matrices in spin space. Under symmetry operations, the irreducible pairing matrices transform as:

$$\hat{\Delta}_{A_{1g},B_{2g}} \rightarrow U\hat{\Delta}_{A_{1g},B_{2g}}U^T$$

The $A_{1g}$ pairing matrix is even under both $\hat{\sigma}_d$ and $\hat{\sigma}_v$, and thus also even under $\hat{C}_4$, while the symmetry of the $B_{2g}$ pairing matrix is odd under $\hat{\sigma}_d$ and even under $\hat{\sigma}_v$, and thus odd under $\hat{C}_4$. These symmetry properties exclude the $A_{1g}$ pairing states and leave only the $B_{2g}$ pairing state, which is exactly the nodeless *d*-wave pairing state discussed in our work, as the pairing order parameter consistent with our experiments.

Model calculations on the SJTM pattern of the nodeless *d*-wave pairing state

In order to help visualize the SJTM pattern, we perform direct model calculations based on Ref. (*44*). For simplicity, we work in the limit of negligible interband SOC and interband pairing as revealed by previous studies (*48*, *51*), under which the 2-Fe UC description is equivalent to the 1-Fe UC description. By extending the theoretical model in Ref. (*44*) to the multiband case, we can express $I_c$ as:

$$I_c(\mathbf{r}) \propto \sum_{\mathbf{k},\alpha} \Delta_\alpha(\mathbf{k})^* \gamma_\alpha(\mathbf{k}) |W_{\mathbf{k}}^\alpha(\mathbf{r})|^2$$

where $\alpha = 1,2$ labels the two relevant bands crossing the Fermi level. $W_{\mathbf{k}}^\alpha(\mathbf{r}) = \sum_{\mathbf{R}} e^{i\mathbf{k}\cdot\mathbf{R}} w_{\mathbf{R}}^\alpha(\mathbf{r})$ is the Fourier transform of the Wannier function of the $\alpha$ orbital/band at momentum $\mathbf{k}$. The factor $\gamma_\alpha(\mathbf{k})$ is given by

$$\gamma_\alpha(\mathbf{k}) = \int \frac{\mathrm{d}\omega}{2\pi} n_F(\omega) \mathrm{Im}\left[\frac{F_t(\omega)}{(\omega + i0^+)^2 - E_{\mathbf{k}}^{\alpha\,2}}\right]$$

where $n_F(\omega)$ is the Fermi distribution function, $E_{\mathbf{k}}^\alpha$ is the Bogoliubov quasiparticle energy, and $F_t(\omega)$ is the anomalous Green's function of the superconducting tip:

$$F_t(\omega) = -i\mathrm{sgn}(\omega) \frac{\pi N_0 \Delta_t}{\sqrt{(\omega + i0^+)^2 - \Delta_t^2}}$$

where $N_0$ is the tip density of states at the Fermi level and $\Delta_t$ is the *s*-wave gap of the tip. $\gamma_\alpha(\mathbf{k})$ and $W_{\mathbf{k}}^\alpha(\mathbf{r})$ are calculated according to the five-orbital (3*d* orbitals of Fe) tight-binding model provided in Ref. (*55*) for monolayer FeSe/STO. The normal state band dispersions are denoted as $\xi_{\mathbf{k}}^\alpha$ with the orbital weights as $V_{\mathbf{k}}^{n\alpha}$, where $n = xz, yz, x^2\text{-}y^2, xy, z^2$ labels the five 3*d* orbitals. The Fourier transform of the Wannier functions can be calculated as:

$$W_{\mathbf{k}}^\alpha(\mathbf{r}) = \sum_n W_{\mathbf{k}}^n(\mathbf{r}) V_{\mathbf{k}}^{n\alpha}$$

$$W_{\mathbf{k}}^n(\mathbf{r}) = \frac{1}{\sqrt{N}} \sum_{\mathbf{R}} e^{i\mathbf{k}\cdot\mathbf{R}} w_{\mathbf{R}}^n(\mathbf{r})$$

where the Wannier functions $w_{\mathbf{R}}^n(\mathbf{r})$ can be approximated by the 3*d* orbitals of a hydrogen-like atom with an effective nuclear charge $Z_{\mathrm{eff}} = 3.7$ (the effective nuclear charge for Fe-3*d* orbitals, (*56*)). The position $\mathbf{r}$ is in the *xy* plane 0.5 nm above the Fe lattice plane. The two pockets at the Fermi level are dominated by the *xz*/*xy* and *yz*/*xy* orbitals (Fig. S12B), respectively (*55*). We take $\Delta_\alpha(\mathbf{k}) \simeq \Delta_0$ for the *xz*/*xy* band ($|V_{\mathbf{k}}^{xz,\alpha}| > |V_{\mathbf{k}}^{yz,\alpha}|$), and $\Delta_\alpha(\mathbf{k}) \simeq -\Delta_0$ for the *yz*/*xy* band ( $|V_{\mathbf{k}}^{xz,\alpha}| < |V_{\mathbf{k}}^{yz,\alpha}|$ ). The factor $\gamma_\alpha(\mathbf{k})$ can be calculated using the Bogoliubov quasiparticle energy $E_{\mathbf{k}}^{\alpha} = \left[\Delta_\alpha(\mathbf{k})^2 + \xi_{\mathbf{k}}^{\alpha 2}\right]^{1/2}$. We take $\Delta_0$ = 10 meV for the nodeless *d*-wave gap amplitude in the sample and $\Delta_t$ = 1 meV for the *s*-wave tip gap in the calculation. The calculated Josephson critical current $I_c(\mathbf{r})$ is shown in Fig. S14A, which is normalized with respect to its magnitude above the midpoints between the nearest-neighbor Fe sites. Figure S14B shows the corresponding $I_c^2(\mathbf{r})$ map, which is proportional to the Josephson tunneling conductance $g_J(\mathbf{r})$. The model calculation results are very consistent with our experimentally measured $g_J(\mathbf{r})$ map shown in Fig. 3B, supporting our conclusion for the nodeless *d*-wave pairing symmetry in monolayer Fe(Te,Se)/STO.

To show the contributions to the Josephson tunneling from different pockets separately, we also calculated $I_c^+(\mathbf{r})$ and $I_c^-(\mathbf{r})$ due to the Fermi pockets with pairing function $+\Delta_0$ and $-\Delta_0$, respectively. As shown in Fig. S14D-E, $I_c^+(\mathbf{r})$ and $I_c^-(\mathbf{r})$ are non-zero everywhere in real space and show different spatial distributions. This is consistent with the expectation that these two pockets are delocalized in real space. The coherent superposition of these two channels results in the sub-unit-cell phase structure of the total Josephson critical current $I_c(\mathbf{r})$ (Fig. S14A).

Model calculations on the SJTM pattern of the assumed extended $s_\pm$ pairing state

For the $s_\pm$ pairing scenario, as discussed above, the Fermi pockets with opposite signs of the gap function are not related by any symmetry operation. Thus, usually symmetry-protected zero Josephson tunneling lines in real space are not expected, which is consistent with our observation of the relatively uniform $g_J$ in bulk Fe(Te,Se) commonly believed to exhibit the $s_\pm$ gap function. To further examine the extended $s_\pm$ pairing scenario (Fig. S12F), we performed model calculations of the Josephson critical current $I_c$ and $I_c^2(\mathbf{r})$ ($I_c^2$ is proportional to $g_J$) by tuning the relative gap amplitudes on the outer pocket and the inner pocket, which are set to have opposite signs. As shown in Fig. S15, changing the relative gap amplitudes can produce zero $I_c^2(\mathbf{r})$ at some positions, but none of the $I_c^2(\mathbf{r})$ patterns reproduce the zero lines along the diagonal mirror plane with half-unit-cell period observed in our experimental $g_J(\mathbf{r})$ maps. These results further support that the $s_\pm$ pairing state is highly unlikely to produce the zero Josephson tunneling lines shown in Fig. 3B.

Josephson interference and phase-sensitivity in SJTM experiments

In SJTM experiments, the observed Josephson tunneling conductance $g_J$ is proportional to $I_c^2$, which only reflects the magnitude of $I_c$. Therefore, the "phase-sensitivity" in SJTM experiments does not imply the ability to determine the sign of the Josephson critical current $I_c$. Rather, it indicates that the experimental outcomes (the real space patterns of $g_J$) are determined by the phase structure of the superconducting order parameter.

Next, we discuss how the phase-sensitivity can survive considering that the SJTM is in the dynamical Coulomb blockade regime, where the gauge-invariant phase difference $\phi$ across the Josephson junction is strongly fluctuating and not well defined. It seems that due to the strongly fluctuating phase $\phi$, the Josephson tunneling in SJTM should lose its phase-sensitivity. However, in previous experimental studies, the phase-sensitivity is mainly shown in interference experiments, which only requires a stable relative phase between different Josephson tunneling channels. In this case, a stable phase for a specific tunneling channel is not necessary. For example, in the phase-sensitive experiment in cuprates based on a superconducting quantum interference device (SQUID) (*3*), the measured quantity is the dynamical resistance (d$V$/d$I$) of the interferometer slightly above the Josephson critical current. In this case, there is actually a finite bias voltage on the Josephson junctions and the phase of the two Josephson junctions rotate synchronously. But the relative phase between the two Josephson junctions in SQUID remains stable and the interference is protected. Thus, the key to the phase-sensitive experiments is a stable relative phase between the Josephson tunneling channels, rather than a stable phase of a certain Josephson tunneling channel. In our SJTM experiments, although the gauge-invariant phase difference between the tip and the sample is strongly fluctuating, the relative phase between Josephson tunneling channels from the two Fermi pockets is stable, which is protected by the *d*-wave pairing symmetry. The symmetry-protected interference of Josephson tunneling channels provides the partial phase-sensitivity explained below.

The "phase-sensitivity" could be a general concept indicating the experimental outcomes are determined by the phase structure of the superconducting order parameter. For example, the quasiparticle interference experiments have been regarded as phase-sensitive in a broad sense because the scattering processes are sensitive to the sign of the order parameter (*57*, *58*). Next, we discuss to what extent the SJTM results shown in our results are phase-sensitive by comparing our work with the seminal phase-sensitive experiment in cuprates based on SQUID (*3*). In both works, multiple coherent Josephson tunneling channels and their interference are used to show the sign-change of the superconducting order parameter. The measured physical quantities ($g_J$ in our work and dynamic resistance $dV/dI$ in the SQUID work) only reflect the magnitude of the Josephson critical current $I_c$ and are not sensitive to its sign. The interference of coherent channels is adjusted by certain tunable degree of freedom (spatial position in our work and magnetic flux in the SQUID work). In the SQUID work, the two coherent channels are two Josephson junctions in a SQUID loop. The Josephson critical current exhibits a local minimum at zero magnetic field, which demonstrates the sign-change of the superconducting order parameter in two perpendicular directions related by four-fold rotation. In our work, the coherent channels originate from different Fermi pockets, related by four-fold rotation or diagonal mirror reflection, enabling orbital dependent pairing order parameters of opposite signs. The Josephson tunneling vanishes along the diagonal mirror lines (the next-nearest-neighbor Fe-Fe lines), suggesting the *d*-wave phase structure of the superconducting order parameter. Thus, in both works, the destructive interference of coherent channels is the key evidence for the *d*-wave symmetry of the order parameter. The most important difference between these two works is that, in our work, changing the tip position can effectively tune the weights of coherent channels, while the interference is always destructive. In contrast, in the SQUID work, the magnetic flux effectively changes the relative phase between these two channels and can drive the interference from destructive to constructive. Thus, the destructive interference of different Josephson tunneling channels in our work provides partial phase-sensitivity. Indeed, our work offers, for the first time, the sub-unit-cell resolution in real space to resolve and detect the *d*-wave superconductivity.

<u>SJTM results on Pb/Si(111) and characterization of the Pb tip</u>

We carry out SJTM measurements on Pb films grown on Si(111) substrate with a Pb tip prepared using the same recipe. Figure S10A shows a typical topographic image of Pb/Si(111) film with atomic resolution. The tunneling spectra along the orange arrow shown in Fig. S10A are obtained using the superconducting Pb tip (Fig. S10C-D) at the base temperature $T$ = 49 mK. The tunneling spectra show clear features at $\Delta_s \approx 1.41$ meV (red dashed line), corresponding to the gap energy of the Pb/Si(111) film. The blue dashed line denotes the voltage corresponding to the sum of the tip and sample gaps $\Delta_t+\Delta_s \approx 2.33$ meV. Thus, the superconducting gap of the Pb tip is $\Delta_t \approx 0.92$ meV. Note that the Josephson tunneling conductance $g_J$ does not show periodic variations at different atomic positions along the orange arrow (Fig. S10D-F). The nearly uniform Josephson tunneling conductance is further confirmed in the spatial map $g_J(\mathbf{r})$ shown in Fig. S10B, which is obtained in the same area as in Fig. S10A. The uniform Josephson tunneling conductance without sub-unit-cell modulations is consistent with the *s*-wave superconductivity of the Pb/Si(111) film.

<u>SJTM results on monolayer FeSe/STO</u>

We carry out SJTM experiments on the monolayer FeSe/STO films. As shown in Fig. S8, the Josephson tunneling conductance shows similar half-unit-cell periodic modulations as in Fe(Te,Se)/STO. The $g_J(r)$ curve observed in the monolayer FeSe/STO shows spatial modulations with the period of half the Se lattice constant $a_{Se}/2$ and the $g_J$ values at the modulation minima are extremely close to zero. The vanishing $g_J$ values (Fig. S8E) further support our argument about the zero Josephson tunneling signal on the zero lines coinciding with the diagonal mirror lines of the lattice.

SJTM results on bulk Fe(Te,Se)

The bulk Fe(Te,Se) exhibits both electron and hole pockets at the Fermi level and is widely recognized as an *s*-wave superconductor with an $s_\pm$ gap function (Fig. S12, D and E). Figure S9A shows a typical topographic image of bulk Fe(Te,Se) with atomic resolution. The Josephson tunneling conductance map $g_J(\mathbf{r})$ is relatively uniform in real space as shown in Fig. S9B. Figure S9C-E displays the tunneling spectra measured along the orange arrow shown in Fig. S9A. The Josephson tunneling conductance is relatively uniform and shows inhomogeneity over length scales much larger than the atomic spacing (Fig. S9F). These results are consistent with previous SJTM studies of bulk Fe(Te,Se) (*37*) and completely different from our results for monolayer Fe(Te,Se) or FeSe on STO and the *d*-wave cases as discussed theoretically above.

SJTM results on $Bi_2Sr_2CaCu_2O_{8+x}$

To further support our experimental technique and theoretical analyses, we also perform SJTM experiments on $Bi_2Sr_2CaCu_2O_{8+x}$, which is a high-$T_c$ cuprate superconductor known to have a *d*-wave pairing symmetry. Figure S11A shows the atomic-resolved topographic image of $Bi_2Sr_2CaCu_2O_{8+x}$ on the BiO surface. The visualized atoms are Bi, which reside right above the Cu atoms in the $CuO_2$ plane underneath. The blue and red balls mark the Cu and O atomic sites in the $CuO_2$ plane, respectively. The Josephson tunneling conductance map (Fig. S11B) shows remarkably similar sub-unit-cell modulations as in the monolayer Fe(Te,Se)/STO. The Josephson tunneling signal is maximized above the O sites between the nearest-neighbor Cu atoms in the $CuO_2$ plane, and almost vanishes above the lines connecting the next-nearest-neighbor Cu atoms, which are the diagonal mirror lines of the lattice. The small but nonzero $g_J$ values on these lines may originate from other tunneling processes, considering that there are residual single-particle states near the nodes in $Bi_2Sr_2CaCu_2O_{8+x}$. As shown in Fig. S11D-E, the tunneling spectra measured along the nearest-neighbor O-O line (orange arrow in Fig. S11C) show a small but nonzero conductance background along with the modulating Josephson tunneling signal (peak structure at zero bias). The extracted Josephson tunneling signal (Fig. S11F) is maximized over the O sites, and vanishes at the midpoints of the nearest O atoms, which are on the next-nearest-neighbor Cu-Cu lines. The experimental results agree with the $I_c^2$ pattern expected from the $d_{x^2-y^2}$-wave pairing symmetry of the cuprate superconductors (*44*), suggesting that our experimental techniques and observations, theoretical analyses, findings and conclusions are consistent, robust, and reliable.

Comparison with previous *c*-axis planar Josephson junction experiments

Covering a *d*-wave superconductor with an *s*-wave layer to form a planar Josephson junction has been used in the early studies of the pairing symmetry in the cuprates (*2*). Small but finite Josephson effect has been observed in various cuprate/Pb junctions (*59–61*), which might originate from spatial inhomogeneities such as dopants, surfaces, interfaces, or steps on a surface. Thus, the macroscopic planar junction experiments could not provide very clear evidence for *d*-wave superconductivity for the cuprates. Moreover, for our monolayer Fe(Te,Se) films grown on STO substrate, such a planar tunnel junction experiment is very challenging.

The monolayer Fe(Te,Se) film is only one unit-cell thick (0.58 nm), which is highly sensitive to air exposure, making the fabrication of a stable planar junction extremely difficult.

Suppressed Josephson tunneling signal around the defect

The sign-reversal superconductivity is sensitive to both magnetic and nonmagnetic defects. As shown in Fig. S16, the Josephson tunneling signal $g_J(\mathbf{r})$ is suppressed around a defect marked by the red dashed circle, which is consistent with the sign-reversal superconductivity.

Absence of half lattice periodic modulations in other tunneling processes

The differential conductance maps at different voltages $g(\mathbf{r}, V)$ are shown in Fig. S17. Only $g(\mathbf{r}, V)$ maps with $V$ extremely close to zero (-0.1 mV to 0.1 mV) show half lattice periodic modulations. This demonstrates that the half lattice periodic modulations are unique patterns of the Josephson tunneling signal and are absent in other non-phase-sensitive tunneling processes. Therefore, the observed half-unit-cell Josephson tunneling modulation cannot be caused by variations in the junction conductance. Furthermore, we display the tunneling spectra at four representative positions (Fe atom, top Te/Se atom, bottom Te/Se atom, and the midpoint) in Fig. S4. Across these positions, the fully gapped single-particle tunneling conductance spectra are comparable at large bias (Fig. S4A), while the Josephson tunneling conductance spectra around zero bias exhibit pronounced spatial variations (Fig. S4B). This further excludes the possibility that the observed zero lines are related to changes in the junction conductance at different positions.

Discussion on higher harmonic origin of the $g_J(\mathbf{r})$ modulation

The Josephson tunneling signal $g_J(\mathbf{r})$ exhibits $2\mathbf{Q}_{a,b}$ modulations, which may cause a question that if the $2\mathbf{Q}_{a,b}$ modulations in $g_J(\mathbf{r})$ originate from the higher harmonic modulations of the lattice wavevector $\mathbf{Q}_{a,b}$. This can be easily addressed because the real space $g_J(\mathbf{r})$ map (Fig. 3B and Fig. S6B) clearly shows that the $2\mathbf{Q}_{a,b}$ modulations are dominant. This is inconsistent with the characteristic of higher harmonic modulations which should be much weaker than the basic frequency modulations. Besides, the higher harmonic scenario cannot explain the "zero lines" observed in our SJTM results.

Possible origins of the finite $g_J(\mathbf{r})$ values on "zero lines"

It is noticed that in Fig. 2D, the $g_J$ is slightly larger than zero at the modulation minima, which could have multiple origins. First, there are other tunneling processes besides Josephson tunneling which also contribute to the zero-bias tunneling conductance. The $I$-$V$ curves simultaneously measured with the d$I$/d$V$ curves in Fig. 2B are shown in Fig. S18A (black curves), which exhibit Josephson tunneling current coexisting with a linear background (red curve). The tunneling conductance contributed from the Josephson tunneling shows a dip-peak-dip structure in a very small range around zero bias (typically within ±0.2 mV at ultralow temperatures), while other tunneling processes are rather uniform within such a small energy range and contribute to roughly linear background in $I$-$V$ curves. The $I$-$V$ background shows a positive slope which originates from other tunneling processes and gives rise to a finite conductance $g_0 \approx 0.064$ μS. After subtracting the conductance from the background, the modulation minima are extremely close to zero as shown in Fig. S18B. Moreover, in SJTM measurements, the measurement positions are discrete and inevitably deviate from the precise "zero lines", leading to a situation where the Josephson tunneling processes cannot cancel out perfectly and exhibit residual non-zero Josephson current at the modulation minima. Such "zero lines" are further supported by the SJTM results in the monolayer FeSe/STO, where the $g_J$ values at modulation minima are extremely close to zero (Fig. S8E).

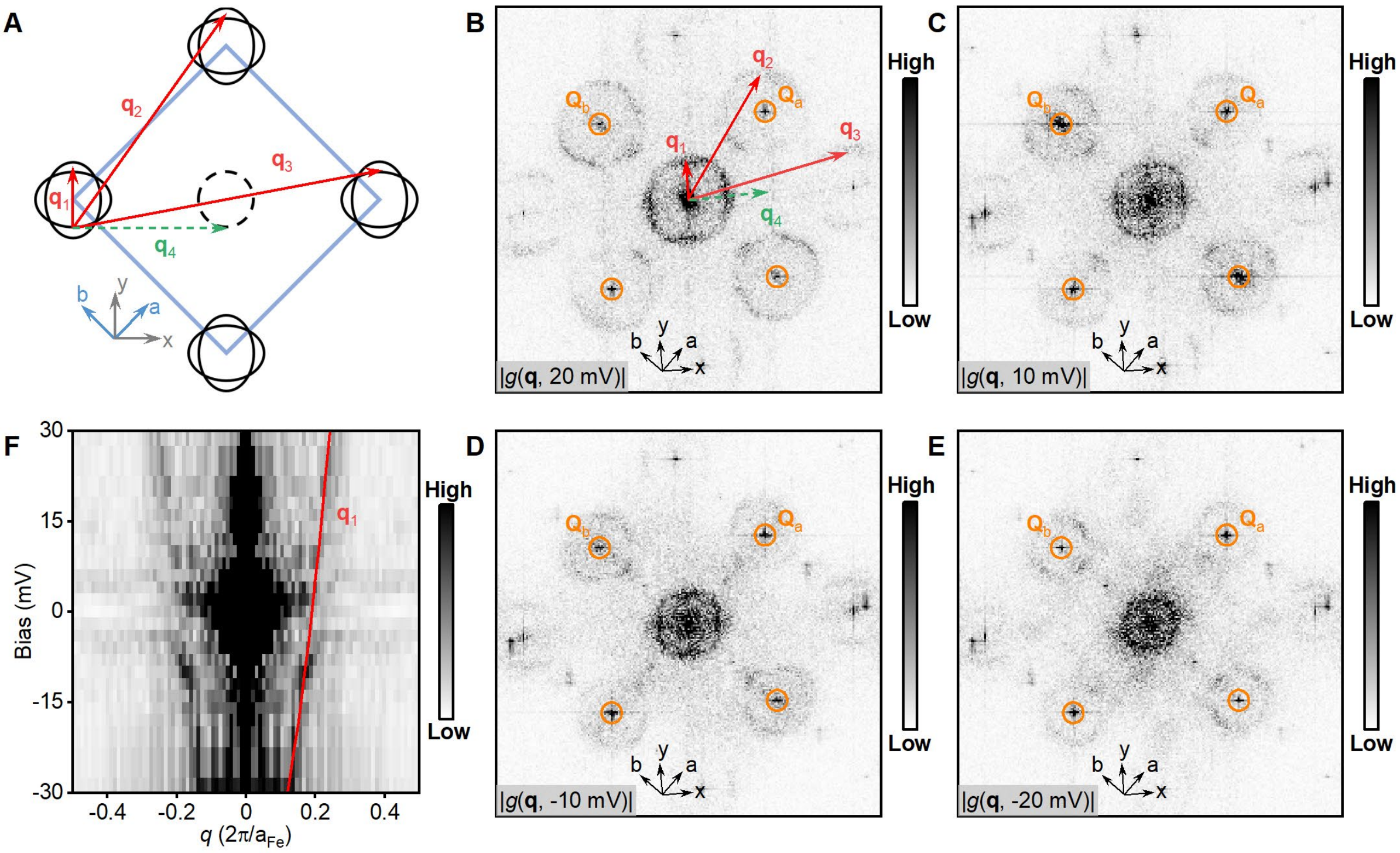

**Fig. S1. Quasiparticle interference results on monolayer Fe(Te,Se)/STO film.** (**A**) Schematic of the Fermi surface of monolayer Fe(Te,Se)/STO in the 2-Fe Brillouin zone. The QPI scattering wavevectors ($\mathbf{q}_{1,2,3}$) connecting electron pockets are marked by red arrows. The QPI scattering wavevector $\mathbf{q}_4$ (dashed green arrow) connects the electron pocket and the absent hole pocket (dashed circle around Γ point). (**B**-**E**) Quasiparticle interference patterns measured at 20 mV, 10 mV, -10 mV, -20 mV. QPI scattering wavevectors are observed at $\mathbf{q}_{1,2,3}$ that correspond to those shown in (A). The $\mathbf{q}_4$ scattering is absent. (**F**) Linecuts of QPI patterns along the $\mathbf{q}_1$ direction in (B) from -30 mV to 30 mV, clearly revealing an electron-type dispersion. The absence of $\mathbf{q}_4$ scattering and the electron-type dispersion of $\mathbf{q}_1$ demonstrate that the Fermi surface only has electron pockets. The QPI results are measured with a metallic PtIr tip. Measurement conditions: $T = 4.3$ K, $V_s = 40$ mV, $I_s = 500$ pA, $V_{mod} = 0.8$ mV.

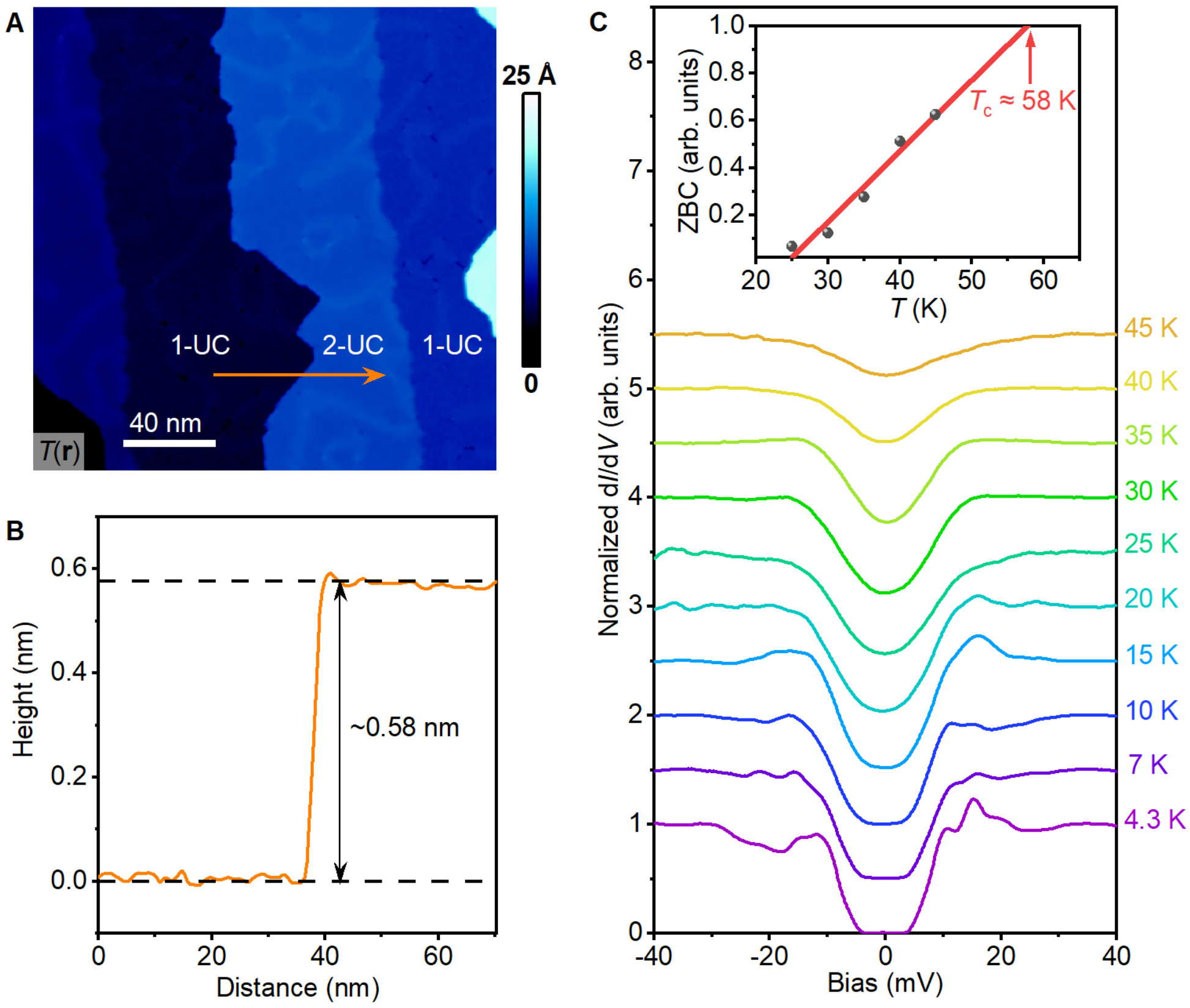

**Fig. S2. More information about the 1-UC Fe(Te,Se)/STO film.** (**A**) A large-scale topographic image of the Fe(Te,Se)/STO film with terrace of STO substrate and partially grown 2nd-UC Fe(Te,Se) ($200\times200$ nm$^2$, $V_s = 1$ V, $I_s = 100$ pA). The right 1-UC area is grown on a higher terrace and is around 0.4 nm (the height of one STO unit cell) higher than the left 1-UC area. (**B**) The altitude line profile taken along the orange arrow in (A). The thickness of the 2nd-UC $FeTe_xSe_{1-x}$ is around 0.58 nm, corresponding to $x \approx 0.3$. (**C**) The temperature dependence of the normalized tunneling spectra, which are obtained by dividing the cubic fitting to the spectra for $|V| \geq 30$ mV (PtIr tip, $V_s = 40$ mV, $I_s = 500$ pA, $V_{mod} = 0.8$ mV). The inset image shows the zero-bias conductance (ZBC) extracted from the normalized tunneling spectra, yielding an extrapolated gap closing $T_c$ of 58 K when the normalized zero-bias conductance is 1.

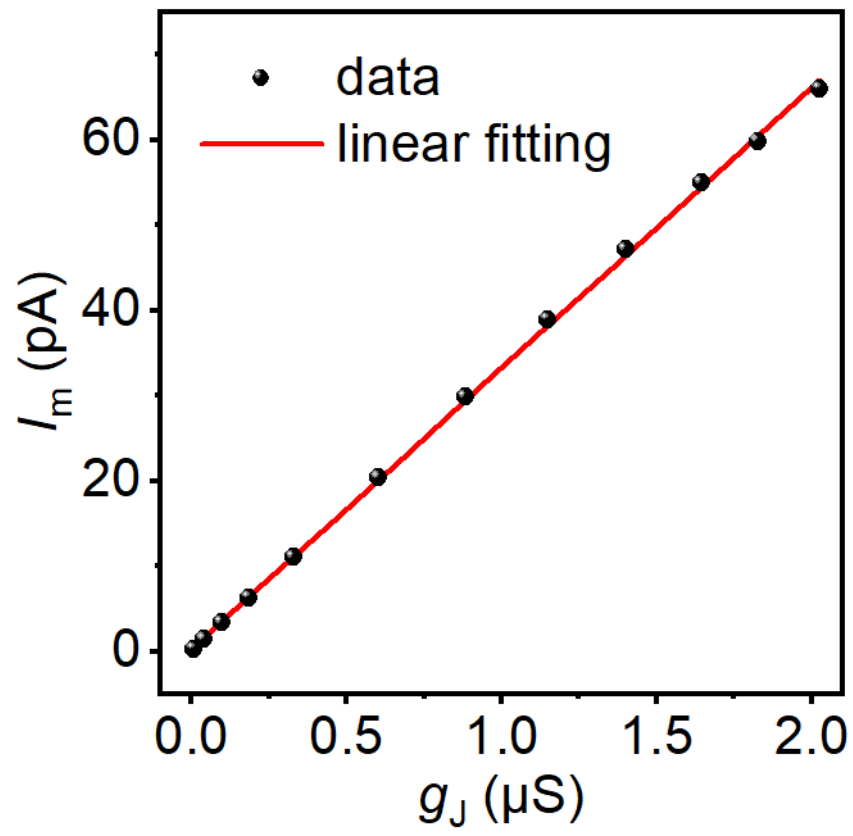

**Fig. S3. Proportional relationship between the $I_m$ and the zero-bias conductance $g_J$, which are proportional to the Josephson critical current squared.**

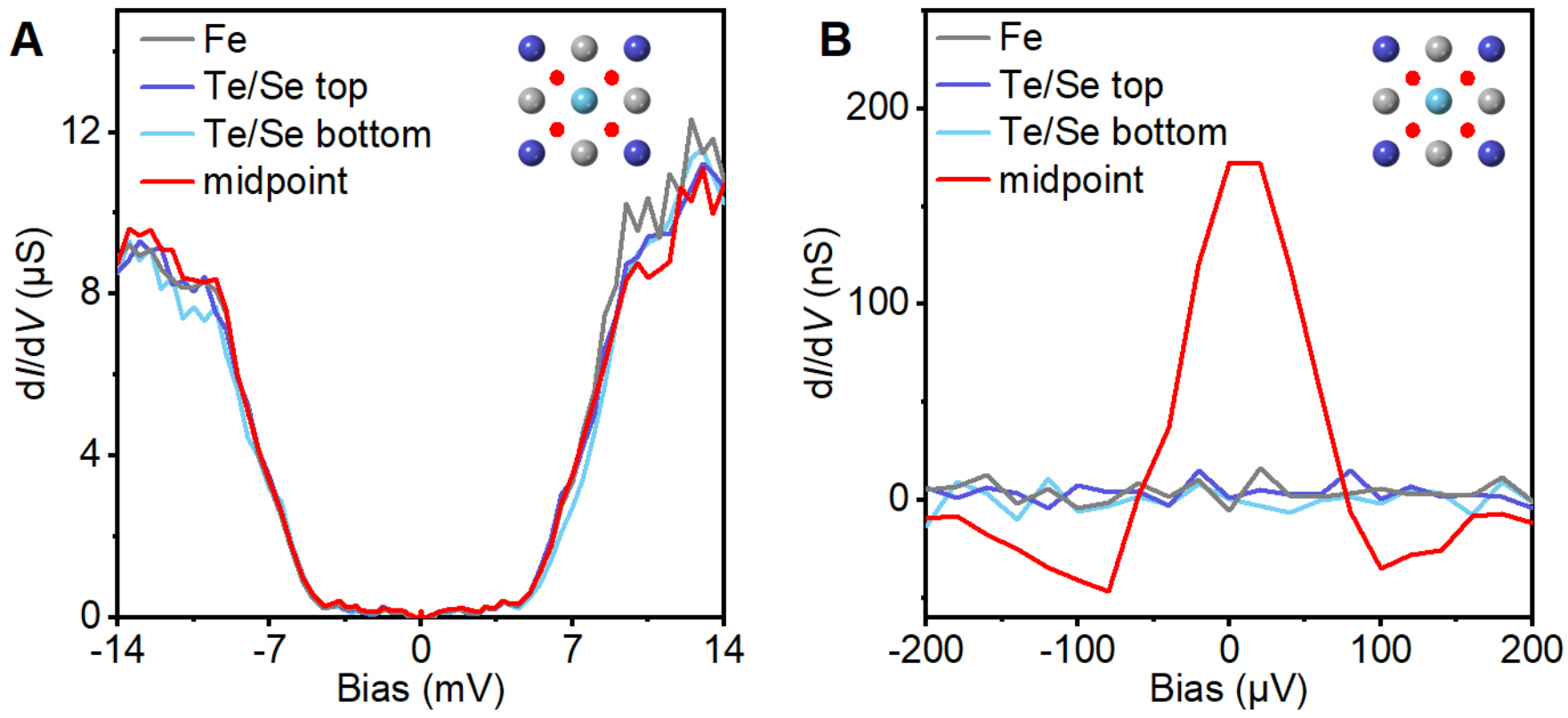

**Fig. S4. Tunneling spectra at different positions.** (**A**) Typical tunneling spectra showing the full gap measured at four different kinds of positions ($T$ = 99 mK, $V_s$ = 14 mV, $I_s$ = 80 nA, $V_{mod}$ = 50 μV). (**B**) Zoom-in spectra around zero bias of (A). At different positions, the single-particle tunneling conductance spectra (A) are comparable while the Josephson tunneling conductance spectra exhibit large variations (B).

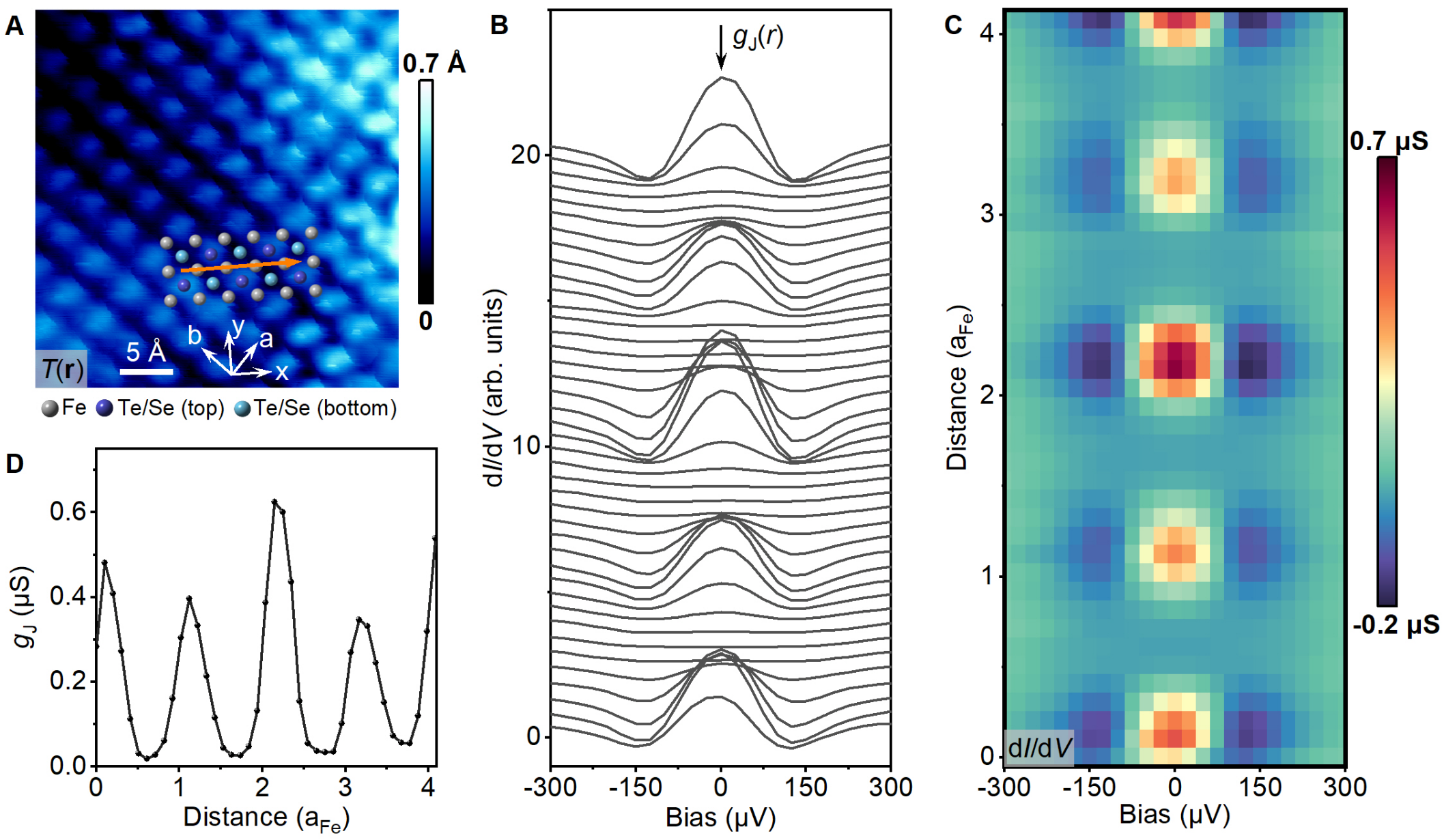

**Fig. S5. Josephson tunneling signal modulation along nearest-neighbor Fe-Fe lines in the 1-UC Fe(Te,Se)/STO film.** (**A**) A topographic image of the 1-UC Fe(Te,Se)/STO film ($3.4\times3.5$ nm$^2$, $V_s$ = 5 mV, $I_s$ = 6 nA). (**B**) Josephson tunneling spectra measured along the orange arrow in (A) ($T$ = 151 mK, $V_s$ = 5 mV, $I_s$ = 6 nA, $V_{mod}$ = 50 μV). The black arrow indicates the Josephson tunneling conductance at zero bias $g_J$. (**C**) The color map of (B), which clearly exhibits the spatially modulated Josephson tunneling signals. (**D**) The extracted $g_J$ along the arrow in (A). The $g_J(r)$ curve shows spatial modulation with the period of the Fe lattice constant $a_{Fe}$ and the $g_J$ values at modulation minima are almost zero. The distances in (C) and (D) are defined relative to the beginning of the arrow in (A).

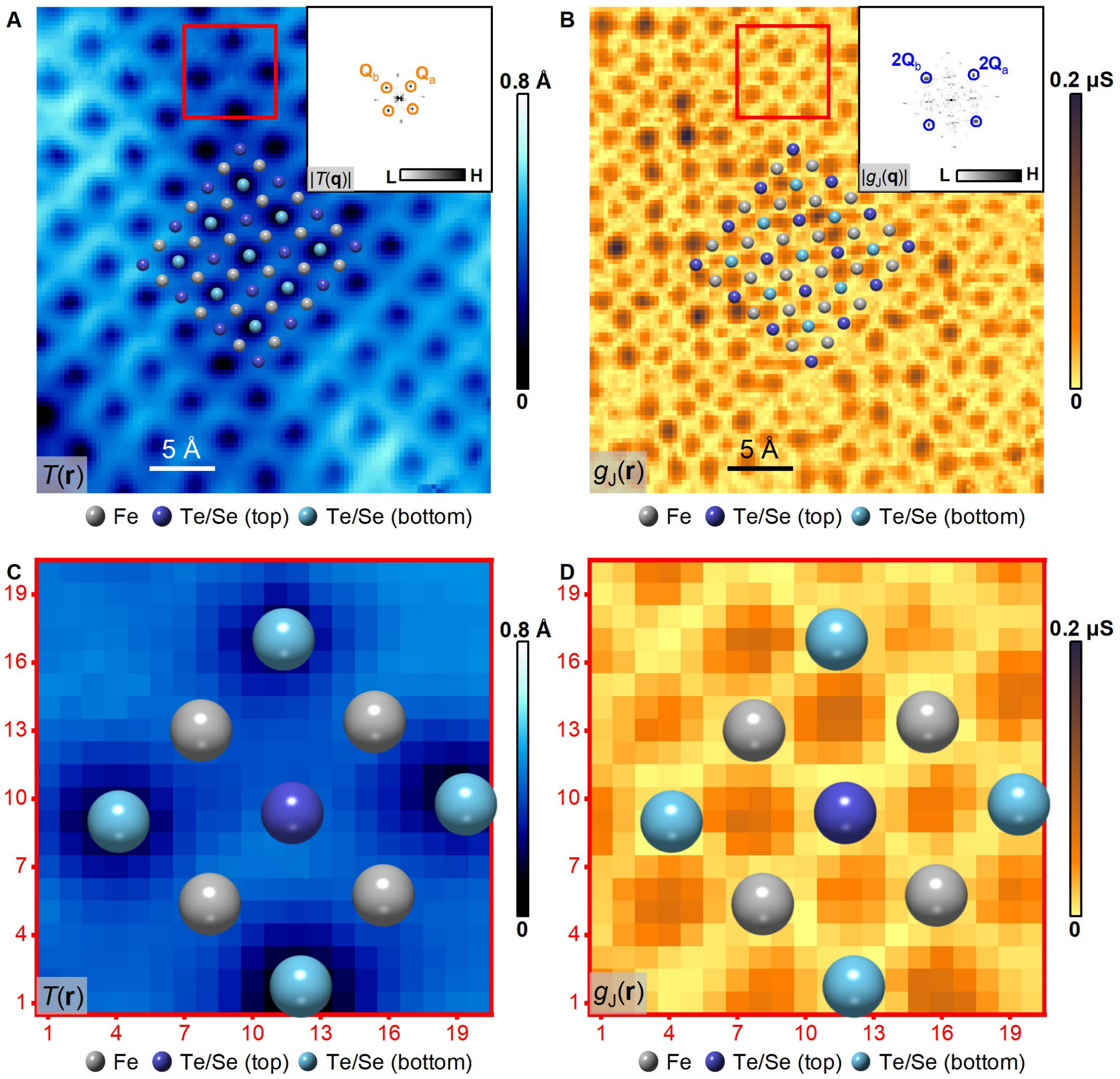

**Fig. S6. Repeatable SJTM results with different setup conditions.** (**A**) A topographic image of the 1-UC Fe(Te,Se)/STO. (**B**) The $g_J(\mathbf{r})$ map measured in the same area as in (A). The topography (A) and $g_J(\mathbf{r})$ map (B) are obtained simultaneously, showing pixel-to-pixel correspondence between these two datasets. Measurement condition: $T = 99$ mK, $V_s = 14$ mV, $I_s = 80$ nA, $V_{mod} = 50$ μV. The sites of Fe and Te/Se atoms within several unit cells are marked by colored balls. The insets of (A) and (B) are the magnitude of the Fourier transform of (A) and (B), respectively. The modulation wavevector peaks at the Bragg points of the topmost Te/Se lattice ($\mathbf{Q}_a$ and $\mathbf{Q}_b$) are denoted by orange circles in the inset of (A). The modulation wavevector peaks at $2\mathbf{Q}_a$ and $2\mathbf{Q}_b$ are denoted by blue circles in the inset of (B). (**C and D**), Zoom-in maps of the red box regions in (A) and (B) respectively, containing 20×20 pixels. The zoom-in maps show clear pixel-to-pixel correspondence between the topography and the $g_J(\mathbf{r})$ map.

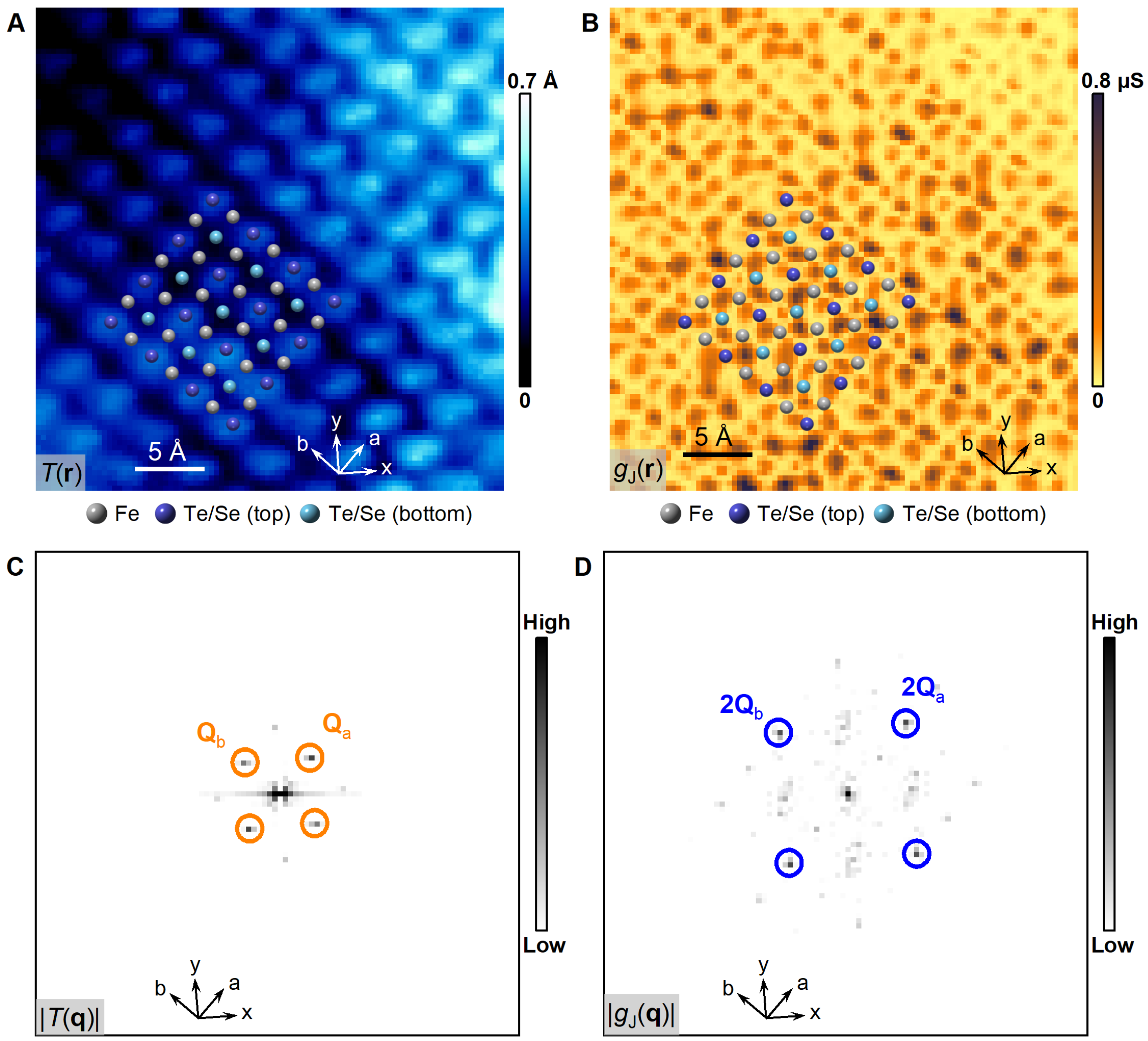

**Fig. S7. Repeatable SJTM results with a different microscopic tip.** (**A**) A topographic image of the 1-UC Fe(Te,Se)/STO. (**B**) The $g_J(\mathbf{r})$ map measured in the same area as in (A). The topography (A) and $g_J(\mathbf{r})$ map (B) are obtained simultaneously, showing pixel-to-pixel correspondence between these two datasets. Measurement condition: $T = 145$ mK, $V_s = 5$ mV, $I_s = 6$ nA, $V_{mod} = 50$ μV. The sites of Fe and Te/Se atoms within several unit cells are marked by colored balls. (**C** and **D**) The magnitude of the Fourier transform of (A) and (B), respectively. The modulation wavevector peaks at the Bragg points of the topmost Te/Se lattice ($\mathbf{Q}_a$ and $\mathbf{Q}_b$) are denoted by orange circles in (C). The modulation wavevector peaks at $2\mathbf{Q}_a$ and $2\mathbf{Q}_b$ are denoted by blue circles in (D).

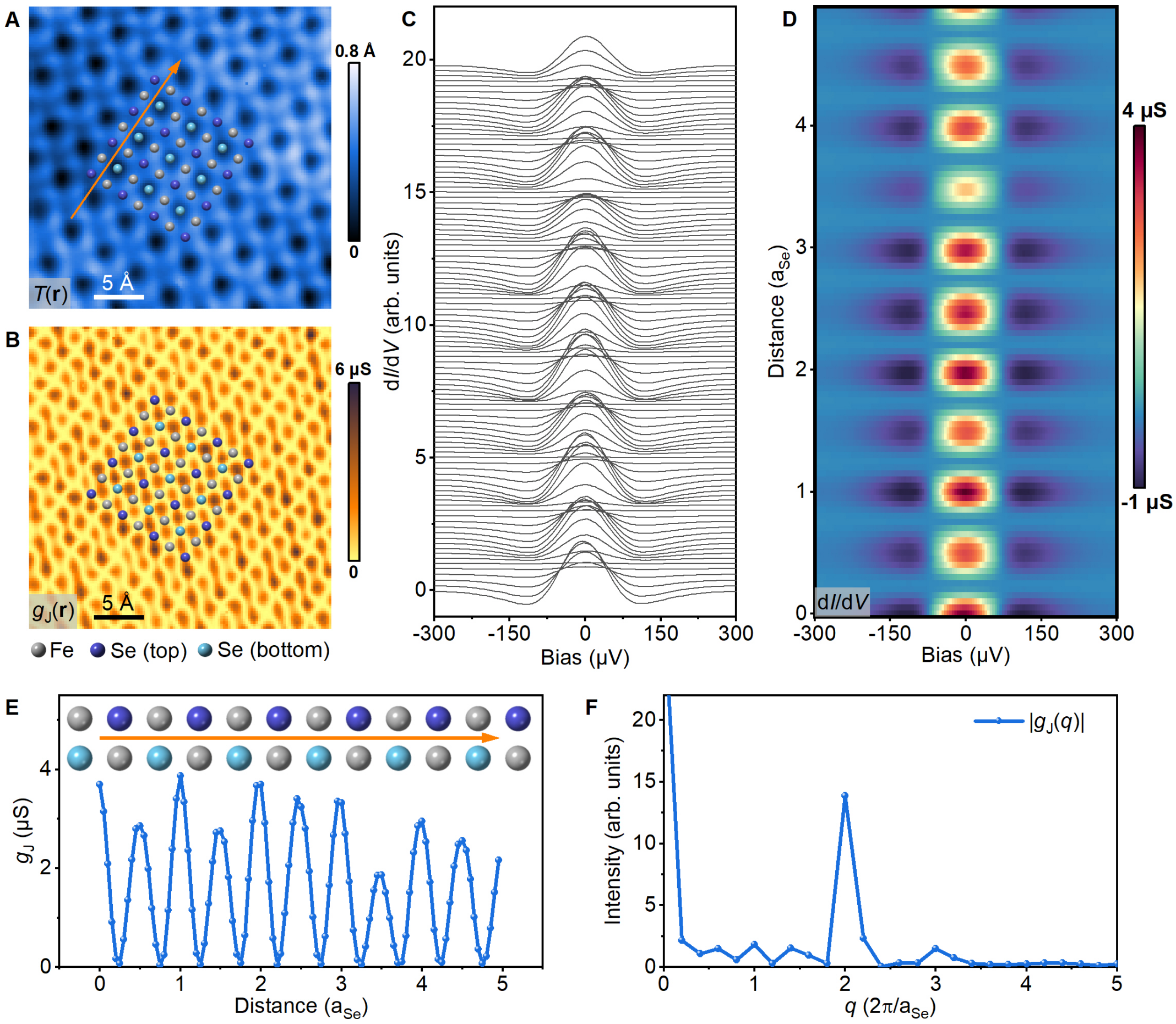

**Fig. S8. SJTM results on monolayer FeSe/STO film.** (**A**) An atomically resolved topography of the monolayer FeSe/STO film (3.0×3.0 $nm^2$). (**B**) The $g_J$(**r**) map measured in the same area as in (A), which shows half lattice periodic modulations. (**C**) The tunneling spectra measured along the orange arrow in (A). (**D**) The color map of (C), which clearly exhibits the spatially modulated Josephson tunneling signals. (**E**) The extracted $g_J$ along the arrow in (A). The $g_J(r)$ curve shows spatial modulations with the period of half the Se lattice constant $a_{Se}/2$ and the $g_J$ values are extremely close to zero at the modulation minima. (**F**) The magnitude of the Fourier transform of (E), showing a strong peak at $q = 4\pi/a_{Se}$. Measurement conditions: $T = 50$ mK, $V_s = 5$ mV, $I_s = 6$ nA, $V_{mod} = 50$ μV.

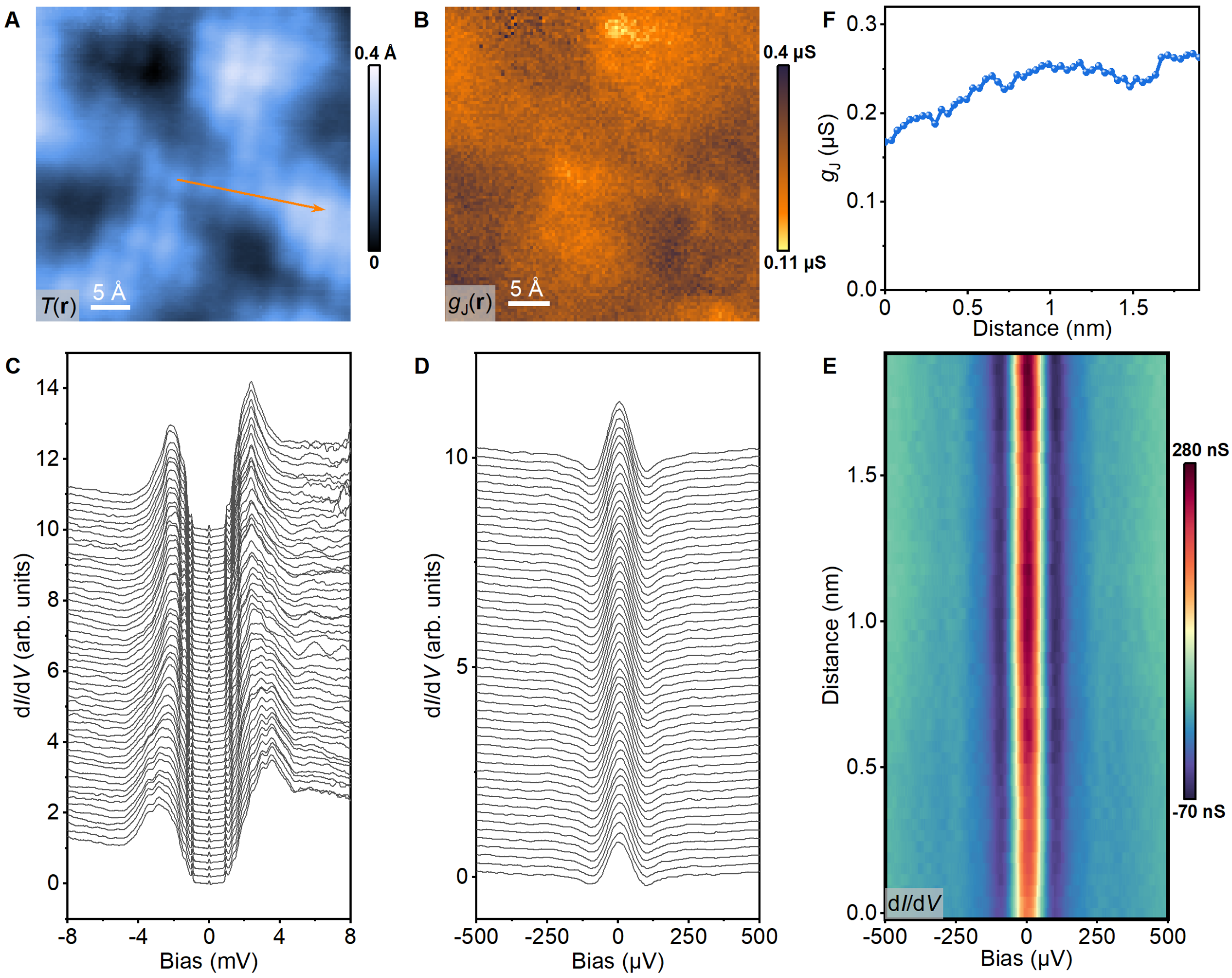

**Fig. S9. SJTM results on bulk Fe(Te,Se).** (**A**) An atomically resolved topography of the bulk Fe(Te,Se) (4.0×4.0 $nm^2$). (**B**) The $g_J$(**r**) map measured in the same area as in (A), which is relatively uniform. (**C**) The tunneling spectra measured along the orange arrow in (A). (**D**) Zoom-in spectra around zero bias of (C). (**E**) The color map of (D). (**F**) The extracted $g_J$ along the arrow in (A). The Josephson tunneling conductance is relatively uniform and shows inhomogeneity over length scales much larger than the atomic spacing, which are consistent with the previous report (*37*) and totally different with our observations in the monolayer Fe(Te,Se) or FeSe on STO. Measurement conditions: $T = 48$ mK, $V_s = -8$ mV, $I_s = 20$ nA, $V_{mod} = 50$ μV.

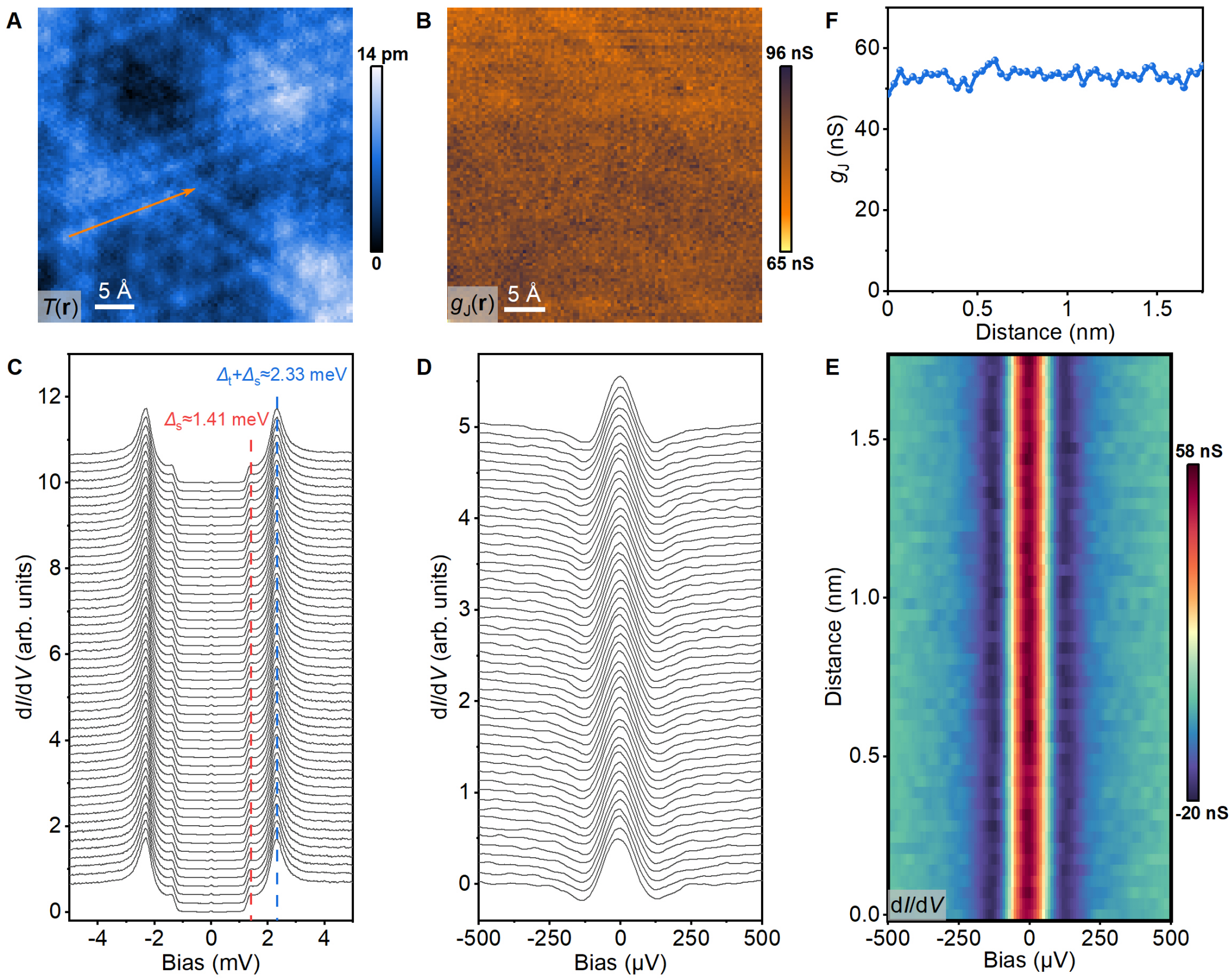

**Fig. S10. Characterization of the superconducting Pb tip on Pb/Si(111) film and SJTM results.** (**A**) An atomically resolved topography of the Pb/Si(111) film (4.0×4.0 nm$^2$). (**B**) The $g_J$(**r**) map measured in the same area as in (A), which is uniform in the field of view. (**C**) The tunneling spectra measured along the orange arrow in (A). The red dashed line denotes the gap energy of the Pb/Si(111) film $\Delta_s \approx 1.41$ meV and the blue dashed line corresponds to the sum of the tip and sample gaps $\Delta_t+\Delta_s \approx 2.33$ meV. (**D**) Zoom-in spectra around zero bias of (C). (**E**) The color map of (D). (**F**) The extracted $g_J$ along the arrow in (A). The Josephson tunneling conductance is uniform and does not show sub-unit-cell modulations in real space, consistent with the conventional *s*-wave superconductivity. Measurement conditions: $T = 49$ mK, $V_s = 5$ mV, $I_s = 8$ nA, $V_{mod} = 50$ μV.

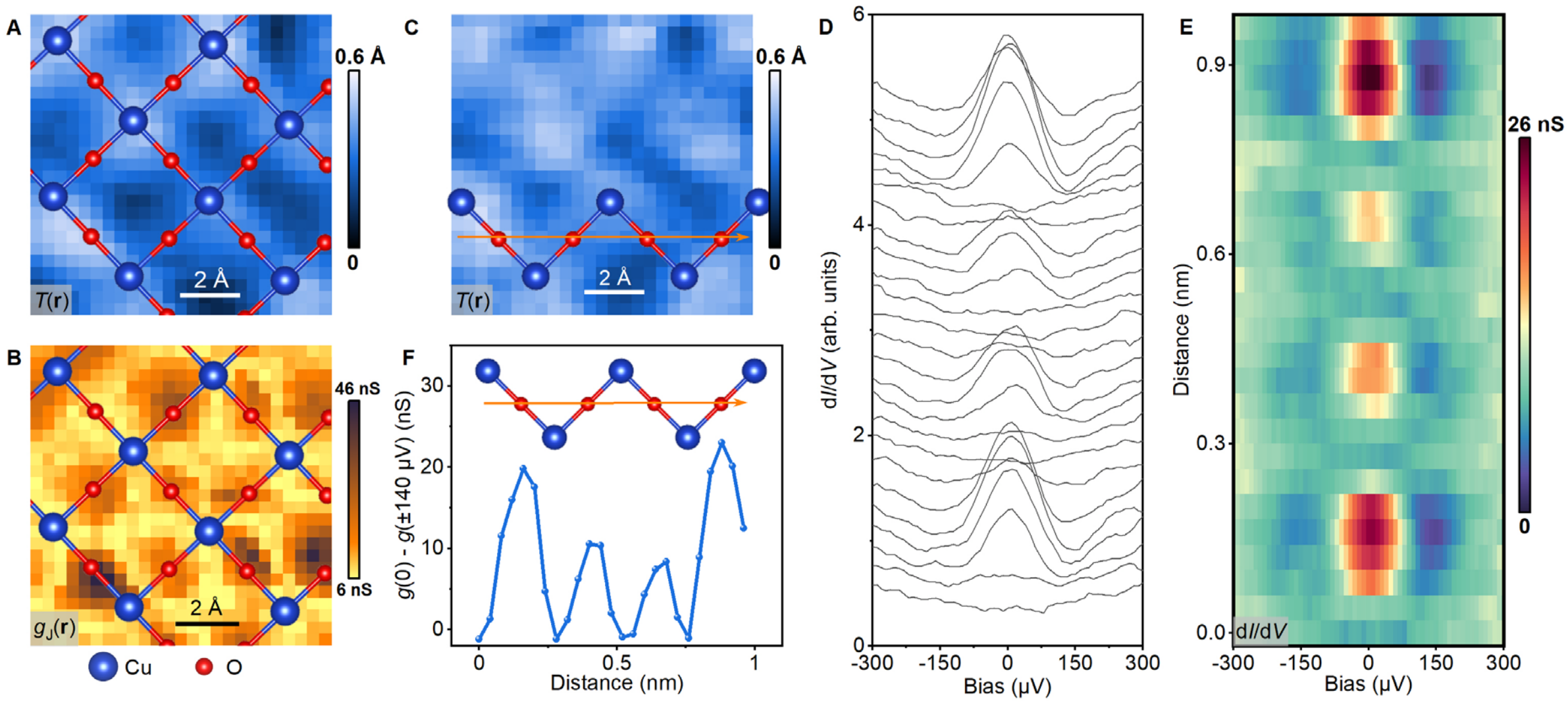

**Fig. S11. SJTM results on $Bi_2Sr_2CaCu_2O_{8+x}$.** (**A**) An atomically resolved topography of the $Bi_2Sr_2CaCu_2O_{8+x}$ on the BiO surface (1.0×1.0 nm$^2$). (**B**) The $g_J(\mathbf{r})$ map measured in the same area as in (A), which shows similar modulations as in the monolayer Fe(Te,Se)/STO. (**C**) Another topographic image of the $Bi_2Sr_2CaCu_2O_{8+x}$ (1.0×1.0 nm$^2$). (**D**) The tunneling spectra measured along the orange arrow in (C). (**E**) The color map of (D). (**F**) The extracted Josephson tunneling signal along the arrow in (C). The Josephson tunneling signal is obtained by subtracting the average tunneling conductance measured at ±140μV from the zero-bias conductance, in order to exclude the contributions from other tunneling processes. The Josephson tunneling signal is maximized at the O sites between the nearest-neighbor Cu atoms in the $CuO_2$ plane, and almost vanishes at the midpoints of the nearest-neighbor O sites, which are on the next-nearest-neighbor Cu-Cu lines. This is consistent with the $I_c^2$ pattern expected from the $d_{x^2-y^2}$-wave pairing symmetry of $Bi_2Sr_2CaCu_2O_{8+x}$. Measurement conditions: $T$ = 50 mK, $V_s$ = -5 mV, $I_s$ = 8 nA, $V_{mod}$ = 50 μV.

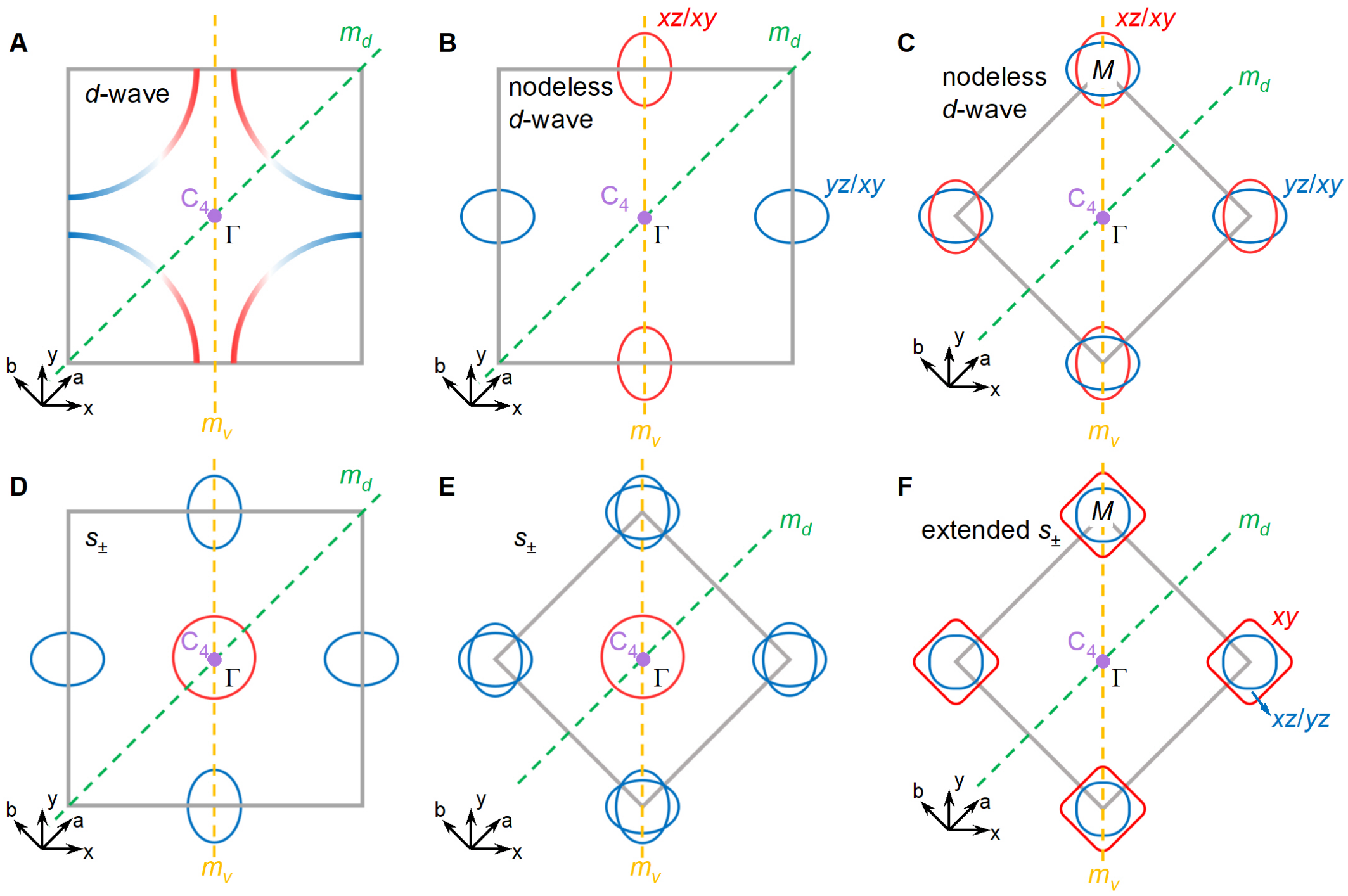

**Fig. S12. Symmetry analysis of different pairing gap functions in k-space.** (**A**) Schematic for the nodal *d*-wave pairing in cuprate superconductors. The symmetry elements, including the mirror planes $m_d$, $m_v$ and the $C_4$ axis are marked by green dashed line, yellow dashed line and purple dot, respectively. Red and blue colors denote opposite signs of the *d*-wave gap function. (**B** and **C**) Schematic for the nodeless *d*-wave pairing in the one-Fe BZ (B) and two-Fe BZ (C) of monolayer Fe(Te,Se)/STO. Red and blue colors denote two Fermi pockets derived from the *xz*/*xy* and *yz*/*xy* orbitals with opposite signs of the gap function. These two Fermi pockets are related by the diagonal mirror reflection $\hat{\sigma}_d$ or the four-fold rotation $\hat{C}_4$. (**D** and **E**) Schematic for the $s_\pm$ pairing in bulk Fe(Te,Se) in the one-Fe BZ (D) and two-Fe BZ (E). The superconducting gap function has opposite signs on the hole pocket around Γ (red) and the electron pockets (blue). Note the *s*-wave pairing symmetry defined by the diagonal mirror reflection $\hat{\sigma}_d$ or the four-fold rotation $\hat{C}_4$. The hole and electron pockets are not related by any symmetry operation. (**F**) Schematic for the extended $s_\pm$ pairing scenario in monolayer Fe(Te,Se)/STO. The superconducting order parameter has *s*-wave symmetry with the gap function having opposite signs on the *xy* pocket (red) and the *xz*/*yz* pocket (blue). The *xy* and *xz*/*yz* pockets are not related by any symmetry operation.

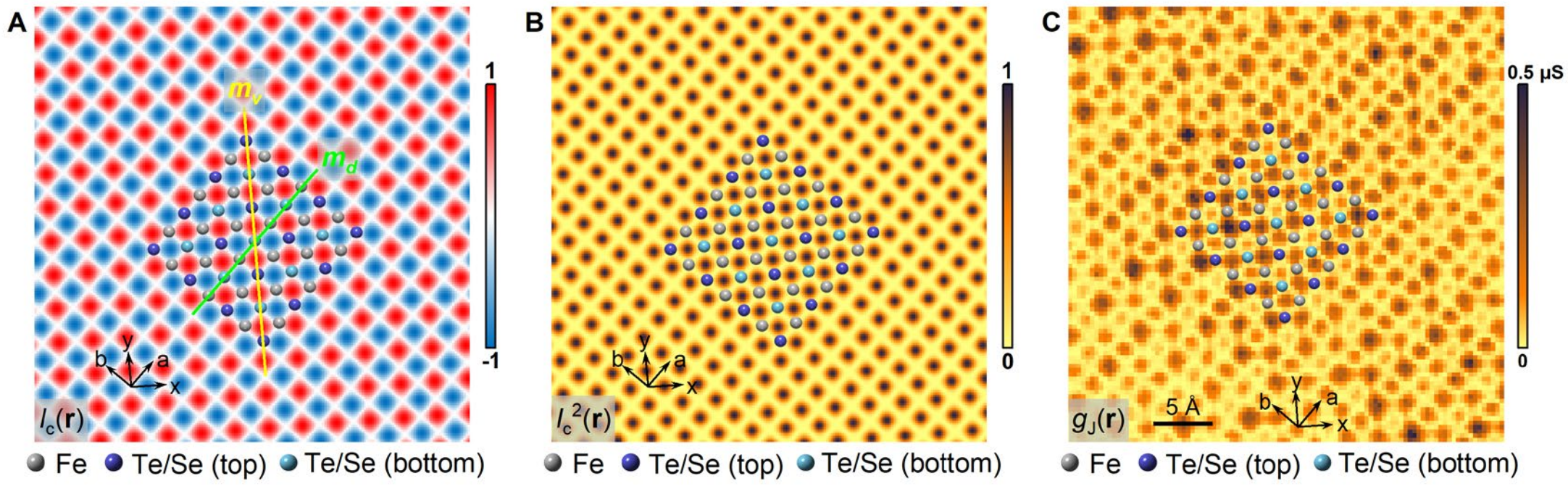

**Fig. S13. Josephson critical current map between an *s*-wave superconducting tip and the monolayer Fe(Te,Se)/STO with nodeless *d*-wave pairing symmetry based on symmetry analyses.** (**A**) $I_c$(**r**) map normalized with respect to its magnitude over the midpoints of nearest-neighbor Fe atoms by symmetry analyses. (**B**) Normalized $I_c^2$(**r**) map by symmetry analyses, which is consistent with the experimental results of $g_J$(**r**) map shown in Fig. 3B. (**C**) The Josephson tunneling conductance $g_J$(**r**) map shown in Fig. 3B ($T$ = 95 mK, $V_s$ = 5 mV, $I_s$ = 2 nA, $V_{mod}$ = 50 μV).

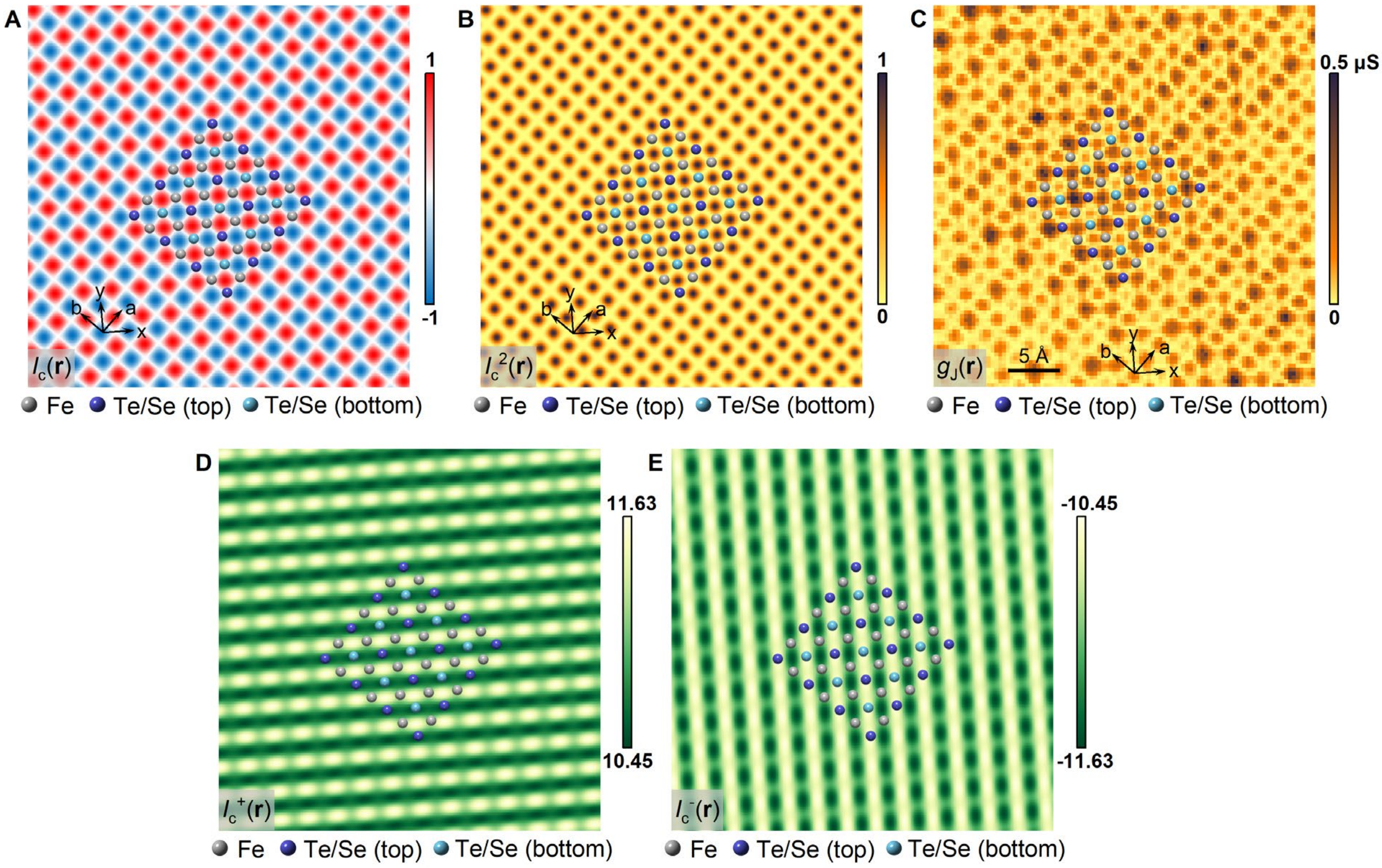

**Fig. S14. Josephson critical current map between an *s*-wave superconducting tip and the monolayer Fe(Te,Se)/STO with nodeless *d*-wave pairing symmetry based on tight-binding model calculations.** (**A**) $I_c(\mathbf{r})$ map normalized with respect to its magnitude over the midpoints of nearest-neighbor Fe atoms. (**B**) Normalized $I_c^2(\mathbf{r})$ map, which is consistent with the experimental results shown in Fig. 3B. (**C**) The Josephson tunneling conductance $g_J(\mathbf{r})$ map shown in Fig. 3B ($T = 95$ mK, $V_s = 5$ mV, $I_s = 2$ nA, $V_{mod} = 50$ μV). (**D** and **E**) The Josephson critical currents $I_c^+(\mathbf{r})$ (D) and $I_c^-(\mathbf{r})$ (E) due to the two Fermi pockets with $+\Delta_0$ and $-\Delta_0$, respectively. The current is also normalized by the magnitude of $I_c$ over the midpoints of the nearest-neighbor Fe atoms. The coherent superposition of these two channels (D and E) results in the sub-unit-cell phase structure of the total Josephson critical current $I_c(\mathbf{r})$ shown in (A).

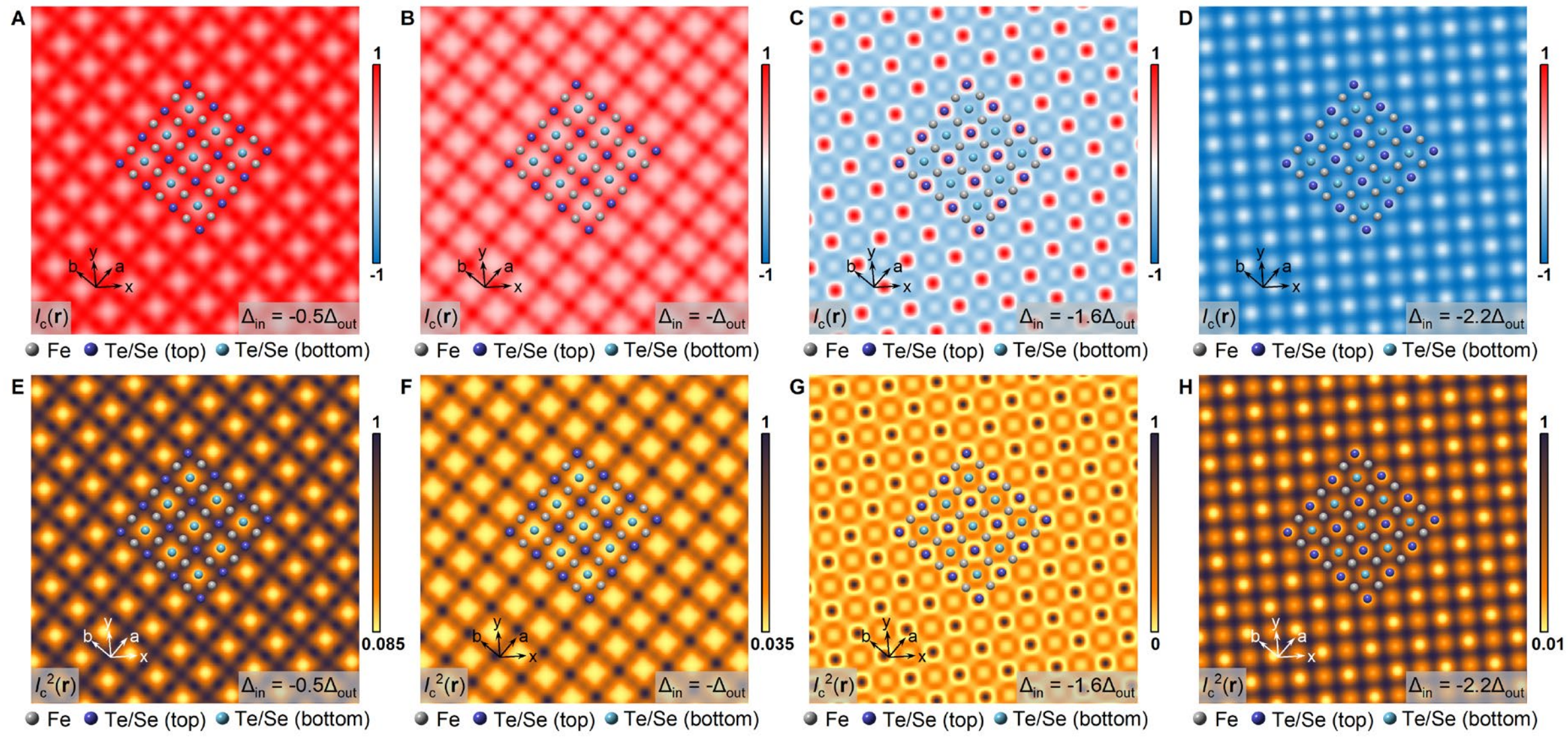

**Fig. S15. Model calculations of Josephson critical current maps between an *s*-wave superconducting tip and the monolayer Fe(Te,Se)/STO with assumed extended $s_{\pm}$ pairing symmetry.** (**A**-**D**) $I_c(\mathbf{r})$ maps normalized with respect to the maximum of $|I_c(\mathbf{r})|$, with different relative gap amplitude of the inner ($\Delta_{in}$) and outer ($\Delta_{out}$) pockets. (**E**-**H**) Normalized $I_c^2(\mathbf{r})$ maps corresponding to $I_c(\mathbf{r})$ maps in (A-D). $I_c$ can be zero or close to zero at some positions by tuning the relative gap amplitudes, but these positions do not form zero lines along the diagonal mirror with half-unit-cell period as observed in our SJTM results.

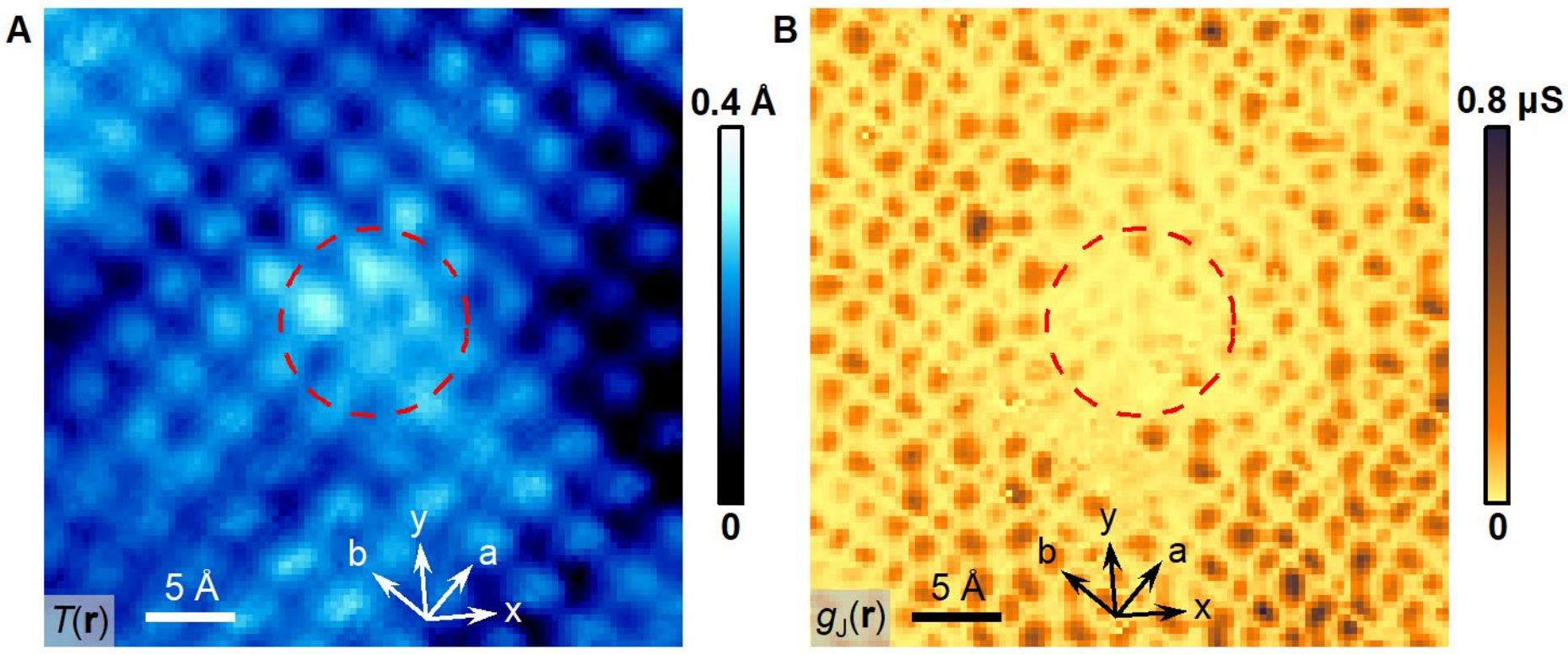

**Fig. S16. Suppression of Josephson tunneling signal near a defect.** (**A**) A topographic image containing a defect circled by red dashed circle. (**B**) The $g_J(\mathbf{r})$ map measured in the same area as in (A). The defect position is marked by red dashed circle. The topography (A) and $g_J(\mathbf{r})$ map (B) are obtained simultaneously, showing pixel-to-pixel correspondence between these two datasets. Measurement condition: $T = 143$ mK, $V_s = 5$ mV, $I_s = 6$ nA, $V_{mod} = 50$ µV. The Josephson tunneling signal is suppressed around the defect.

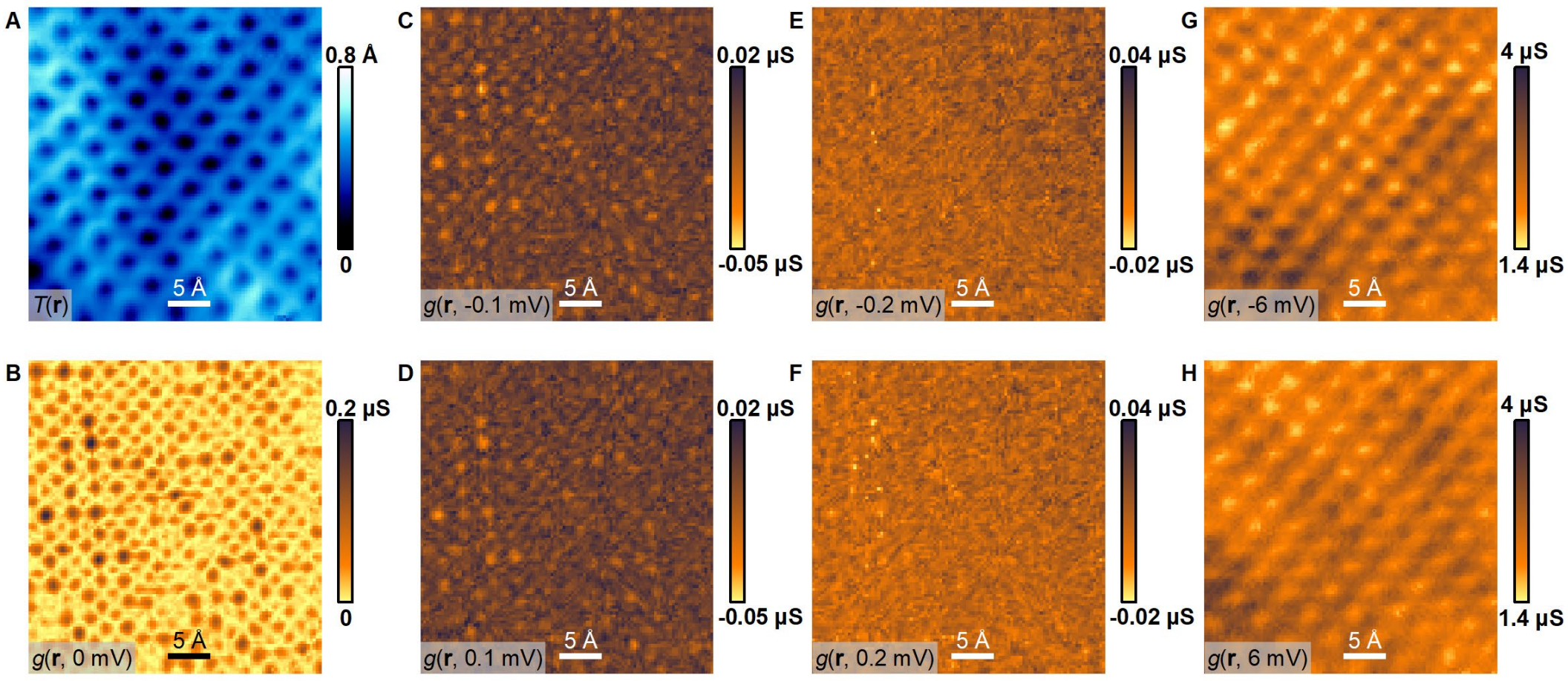

**Fig. S17. Tunneling conductance maps *g*(r, *V*) at different bias voltages.** (**A**) The topographic image of the 1-UC Fe(Te,Se)/STO shown in Fig. S6A. (**B-H**), The *g*(**r**, *V*) maps obtained in the same area as in (A). The topography (A) and *g*(**r**, *V*) maps (B-H) are obtained simultaneously, showing pixel-to-pixel correspondence between these datasets. Only *g*(**r**, *V*) maps with *V* extremely close to zero (-0.1 mV to 0.1 mV) show $2\mathbf{Q}_{a,b}$ modulations. This demonstrates that the $2\mathbf{Q}_{a,b}$ modulations are unique patterns of the Josephson tunneling signal and are absent in other non-phase-sensitive tunneling processes. Measurement condition: $T$ = 99 mK, $V_s$ = 14 mV, $I_s$ = 80 nA, $V_{mod}$ = 50 μV.

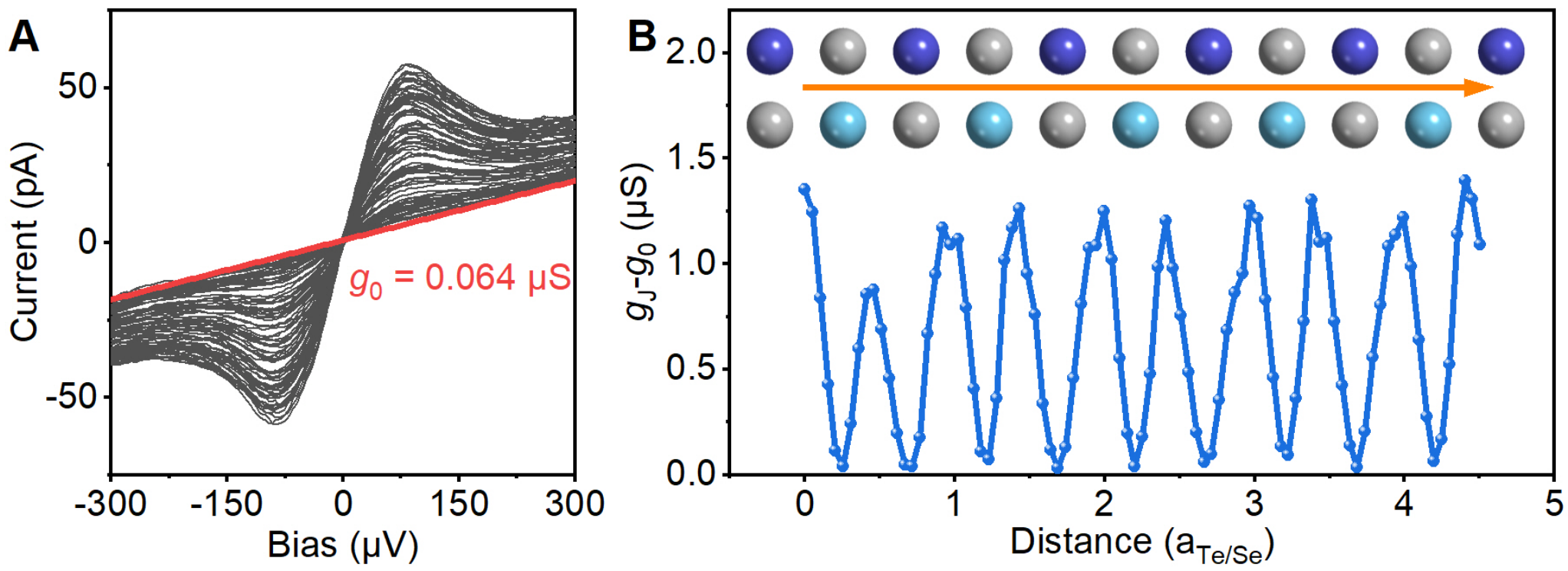

**Fig. S18. Finite conductance contributed by other tunneling processes.** (**A**) Current-voltage ($I$-$V$) curves (black) simultaneously measured with Fig. 2B ($T$ = 70 mK, $V_s$ = 5 mV, $I_s$ = 4 nA, $V_{mod}$ = 50 μV) at different positions. The red curve shows the background originating from other tunneling processes with a tunneling conductance of $g_0 \approx 0.064$ μS. (**B**) The $g_J$-$g_0$ modulation obtained by subtracting $g_0$ from the data in Fig. 2D. The modulation minima are extremely close to zero.